\documentclass[fleqn,usenatbib]{mnras}

\usepackage{newtxtext,newtxmath}

\usepackage[T1]{fontenc}

\DeclareRobustCommand{\VAN}[3]{#2}
\let\VANthebibliography\thebibliography
\def\thebibliography{\DeclareRobustCommand{\VAN}[3]{##3}\VANthebibliography}

\usepackage{graphicx}	
\usepackage{amsmath}	
\usepackage{siunitx}[=v2]

\newcommand{\vvec}{\ensuremath{\mathbf{v}}}
\newcommand{\vveca}{\ensuremath{{\mathbf{v}_2}}}
\newcommand{\kvec}{\ensuremath{\mathbf{k}}}
\newcommand{\kveca}{\ensuremath{{\mathbf{k}_2}}}
\newcommand{\uvec}{\ensuremath{\mathbf{u}}}
\newcommand{\uveca}{\ensuremath{{\mathbf{u}_2}}}

\title[Three dimensional Doppler tomography]{Three dimensional Doppler tomography}

\author[T.R.\ Marsh]{
T.R.\ Marsh\thanks{E-mail: t.r.marsh@warwick.ac.uk}\\
Department of Physics, University of Warwick, Coventry CV4 7AL, United Kingdom
}

\date{Accepted XXX. Received YYY; in original form ZZZ}

\pubyear{2015}

\begin{document}
\label{firstpage}
\pagerange{\pageref{firstpage}--\pageref{lastpage}}
\maketitle

\begin{abstract}
Doppler tomography is a method to compute the emissivity distribution within the co-rotating frames of binary stars from observations of their emission line profiles at multiple orbital phases. A key assumption of the method as it is usually applied is that all gas flow is parallel to the orbital plane of the binary. In this paper I examine the possibility of lifting this assumption to allow for motion parallel to the orbital "$z$" axis of the binary as well. I show that the problem is best considered in Fourier space, and that line profiles directly constrain the 3D Fourier transform of the 3D Doppler image in velocity space, but only over the 2D surface of a double-cone centred upon the origin, and aligned with the axis reciprocal to the $v_z$ velocity axis. Hence the  full information needed for the recovery of the 3D emissivity distribution is simply not available. Despite this, an inversion method is presented and tested on a number of simulated images. While artefacts resulting from the missing information do appear, the tests suggest that there could be some value in applying 3D Doppler tomography to data from real systems, although considerable care is needed when doing so.
\end{abstract}




\section{Introduction}

Atomic lines from binary stars vary in wavelength as the binary orbit progresses. Any component that is fixed in the frame of the binary does not appear invariant to us, but instead it executes a sinusoid in terms of its radial velocity with time (for assumed circular orbits). Such  sinusoids were seen long ago as "S-waves" in trailed photographic plate spectra \citep{1962ApJ...136..312K}. If there are many different components, each executing its own sinusoid with its own amplitude and phase, the blended result of overlapping sinusoids can be hard to interpret in terms of the pattern of emissivity in the frame of the binary. Very often in fact, the emissivity has the form of a smooth distribution as opposed to a discrete set of point sources, and a simplistic component--source association becomes impossible.

It was to address this problem in the context of accreting white dwarfs in binaries, which feature line emission from extended structures taking the form of discs and streams of material as well as from their component stars, that the method known as Doppler tomography was developed \citep{1988MNRAS.235..269M}. In essence Doppler tomography seeks to find the pattern of emissivity in the frame of the binary that matches an observed set of line profiles at multiple orbital phases. Doppler tomography was so named because the relationship between the emissivity pattern in the binary and the line profiles as a function of orbital phase is mathematically very similar to the relationship between structures inside the human body and medical x-ray images as a function of the projection angle around the body. The analogy between the two cases is closest if the Doppler images are considered to be a function of velocity since this avoids the often uncertain and potentially multi-valued and therefore indeterminate translation between spatial position and velocity within a binary system. Thus it is that Doppler images are almost invariably presented in two dimensional velocity space, reflecting the motions parallel to the orbital plane of the binary.

In 2D velocity coordinates, the line profiles are a collapse or projection of the
Doppler image along a straight-line direction defined by orbital phase \citep[in contrast to the curved lines that characterise line profile formation from an accretion disc in spatial coordinates for instance,][]{1972ApJ...171..549H,1981AcA....31..395S,1986MNRAS.218..761H}. The projection integrates over the orbital-phase dependent direction to convert the 2D image into a 1D line profile. A set of such projections at all angles (i.e. all orbital phases) can be inverted to obtain the 2D velocity-space image through a process called a Radon transform \citep{1986LNP...266....1H,1988MNRAS.235..269M,
2001LNP...573....1M}, although practical implementations need to account for missing phases and finite spectral and temporal resolution which do not feature in the Radon transform.

Line profiles as a function of orbital phase are two-dimensional in nature, as are Doppler images, so it is perhaps no surprise that the inversion is usually
a well-constrained problem, even in the face of the practical details of noise, resolution and phase coverage. However, some of the most striking results of Doppler tomography have come from its application to the magnetic accreting white dwarf systems known as "polars" \citep[and also as AM~Her stars after the prototype system][]{1977ApJ...212L.125T}.
Doppler images of polars have revealed the ballistic part of the mass transfer stream, along with emission from the white dwarf's magnetosphere once the accreting gas has locked onto the magnetic field lines and as it accelerates towards the white dwarf \citep{1997A&A...319..894S,1999MNRAS.304..145H,2000MNRAS.313..533S}. 
Magnetically confined accretion in polars could well involve significant motion out of the orbital plane, violating a fundamental assumption of Doppler tomography \citep{2001LNP...573....1M}.
The application of Doppler tomography to polars raises two obvious questions: what is the effect of out-of-plane motion upon the images, and can it be accounted for in the process of Doppler tomography?

Out of plane motion corresponds to motion in the $z$ direction, parallel to the orbital axis of the binary. This implies a $v_z$ component in addition to the usual $(v_x,v_y)$ coordinates of Doppler tomography. In other words the question becomes whether it is possible to reconstruct a fully 3D image from line profile data, and, if so, is the reconstruction reliable? Addressing these questions is the subject of this paper.

A first look at 3D recovery was briefly presented in the form of a simulated image of two spots by \citet{2005Ap&SS.296..403M}. \citet{2006ApJ...652.1547A} applied a method for 3D reconstruction related to the CLEAN technique of radio astronomy to derive 3D maps of the Algol system U~CrB. The same authors returned to U~CrB in \citet{2009ApJ...690.1730A} as well as other Algol binaries \citep{2012ApJ...760....8R}. These papers present a rather bewildering variety of components interpreted in terms of stream and disc flows in the orbital plane, jet outflows along the orbital axis, flows along magnetic loops and coronal mass ejections. In this paper, I will show that there are reasons to be cautious when it comes to the reality of such features. Tomography in 3D is found to have a fundamentally different character than the 2D case, meaning that very different emissivity distributions can be found that match the same set of data equally well. On the flip side there can also be instances of the faithful recovery of 3D information. 

I start with a review of the 2D imaging problem. Although this has been covered before \citep{1988MNRAS.235..269M,2001LNP...573....1M}, it is useful here to summarise it in terms as closely connected as possible to the 3D case that will be studied in 
section~\ref{sec:3dlform}; much of the groundwork for the 3D analysis is contained within this summary. I look at the nature of the degeneracy intrinsic to 3D imaging in section~\ref{sec:anything}, and present a method for computing 3D images from line profile data in section~\ref{sec:implement}. I show simulated reconstructions in section~\ref{sec:simulated}, before finishing with a discussion and conclusions.

\section{Line formation in 2D velocity space}

A point in 2D velocity space is labelled by its velocity
$\vveca = (v_x,v_y)$. The $x$ and $y$ velocity components are 
defined relative to an $x$ axis which points from star~1 to star~2 and
a $y$ axis that points in the direction of motion of star~2. The subscript "2" is used to flag that this is a 2D vector; later on, 3D vectors will appear without subscripts.

The line profile $f(V,\phi)$, a function of radial velocity $V$ and orbital phase $\phi$ (measured in terms of orbital cycles), due to image $I_2$ in 2D velocity space is given by
\begin{eqnarray}
f(V,\phi) &=& 
\iint dv_x \, dv_y \, \delta(V-V_R) I_2(v_x,v_y), \\
&=&\int_\vveca d^2\vveca \; \delta(V-V_R) I_2(\vveca), \label{eq:p2form}
\end{eqnarray}
adopting an integrand-goes-last notation to reduce later complexity,
where the second line introduces a compact notation for the double integral, and $V_R$ is the radial velocity given by
\begin{eqnarray}
V_R &=& - v_x \cos (2\pi \phi) + v_y \sin (2\pi \phi), \label{eq:straight}\\
    &=& \uveca . \vveca . 
\end{eqnarray}
Here the vector \uveca\ is defined as
\begin{equation}
    \uveca = (-\cos (2\pi \phi), \sin(2\pi \phi)).
\end{equation}
The form of the expression adopted here for the radial velocity $V_R$
means that $v_x$ and $v_y$ implicitly include a $\sin(i)$
projection factor along the line-of-sight where $i$ is the orbital inclination. The inclusion of the projection factor in this manner is standard practice in 2D Doppler tomography since the projection factor is usually unknown, and it means that the scale of the derived map is directly connected to the radial velocity in the line profiles. This has the consequence that the true velocities in the orbital plane are a factor $1/\sin(i)$ higher than they appear in 2D Doppler maps. It also means of course that we require that $i \neq \ang{0}$.

Eq.~\ref{eq:p2form} expresses a projection operation. The delta function selects all points in the image that have radial velocity $V_R = V$ at orbital phase $\phi$.  Eq.~\ref{eq:straight} means that these points satisfy $\uveca \cdot \vveca = V$, and form a straight
line perpendicular to \uveca\ in velocity coordinates. Considering all possible values of $V$, then a set of parallel straight lines emerges along which one imagines the 2D image is collapsed to form the observed profile, a "projection" in other words. These ideas were illustrated in the first two figures of \citet{1988MNRAS.235..269M}.

Defining Fourier and inverse Fourier transforms by the relations
\begin{eqnarray}
\tilde{f}(k) &=& \int_{-\infty}^{+\infty} dx \, e^{-i2\pi k x} f(x),\\ 
f(x) &=& \int_{-\infty}^{+\infty} dk \, e^{i2\pi k x} \tilde{f}(k),
\end{eqnarray}
adopting the convention that $\tilde{f}$ is the Fourier transform
of $f$, then taking the 2D transform of the image leads to 
\begin{eqnarray}
\tilde{I}_2(\kveca) &=& \int_\vveca d^2\vveca \, e^{- i 2\pi \kveca \cdot \vveca} I_2(\vveca),\\
I_2(\vveca) &=& \int_\kveca d^2\kveca \, e^{i 2\pi \kveca \cdot \vveca} \tilde{I}_2(\kveca), \label{eq:2dFT}
\end{eqnarray}
with $\tilde{I}_2(\kveca)$ the 2D Fourier transform of the 2D velocity space image. The double integrals span all of the velocity \vveca-space, and the reciprocal \kveca-space, although in practice there are limitations due to size and noise which are not a concern at this point. 

Substituting for $I_2(\vveca)$ from Eq.~\ref{eq:2dFT} into Eq.~\ref{eq:p2form} for the line profiles, and switching the order of integration,
\begin{eqnarray}
f(V,\phi) &=&\int_\vveca d^2\vveca \, \delta(V-V_R) \int_\kveca d^2\kveca \, e^{i2\pi \vveca \cdot \kveca} \tilde{I}_2(\kveca), \nonumber\\
&&\\
&=& \int_\kveca d^2\kveca \int_\vveca d^2\vveca \, \delta(V-\uveca\cdot\vveca) e^{i2\pi \kveca\cdot\vveca} \tilde{I}_2(\kveca).\nonumber\\
&&\label{eq:2dinter}
\end{eqnarray}
Substituting for the delta function in the last line using the well-known
relation
\begin{equation}
    \delta(x) = \int ds \, e^{i2\pi s x},
    \end{equation}
and swapping the order of integration once more,
\begin{equation}
f(V,\phi) = \int_s ds \, e^{i2\pi s V} \int_\kveca d^2\kveca \,
\tilde{I}_2(\kveca) \int_\vveca d^2\vveca \, 
e^{i 2\pi (\kveca - s \uveca)\cdot\vveca} . 
\end{equation}
The double integral over \vveca\ reduces to the product of two delta functions, $\delta(k_{vx} - s u_x)$ and $\delta(k_{vy} - s u_y)$, which I write in compact form as $\delta^2(\kveca - s \uveca)$.
(I refer to $k_{vx}$ and $k_{vy}$ rather than $k_x$ and $k_y$, since $\kveca$ is conjugate to velocity space not position space.) Hence 
\begin{eqnarray}
f(V,\phi) &=& \int_s ds \, e^{i2\pi s V} \int_\kveca d^2\kveca \,
\tilde{I}_2(\kveca) \delta^2(\kveca - s\uveca),\nonumber\\
&&\\
&=& \int_s ds \, e^{i2\pi s V} \tilde{I}_2(s \uveca) .
\end{eqnarray}
Finally, taking the Fourier transform over $V$,
\begin{equation}
    \int dV\, e^{-i 2\pi s V} f(V,\phi) = \tilde{I}_2(s\uveca),
\end{equation}
and so
\begin{equation}
    \tilde{f}(s, \phi) = \tilde{I}_2(s \uveca) . \label{eq:2dkey}
\end{equation}
This equation shows that the Fourier transform with respect to radial velocity of the line profile at phase $\phi$ gives us the values of the 2D Fourier transform of 2D Doppler image along a line in \kveca-space given by $\kveca = s \uveca$ for $s = -\infty$ to $\infty$. This is a straight line through the origin in \kveca-space with direction \uveca. This line lies at angle $2\pi \phi$ radians rotated anti-clockwise from the $k_{vx}$ axis. As the binary orbit progresses, the angle increases, and hence we can obtain values of the 2D Fourier transform over the entire $k_{vx}$--$k_{vy}$ plane given a set of line profiles covering a binary orbit (or even just half a binary orbit indeed). The image that we are after then follows from Eq.~\ref{eq:2dFT}.

To see explicitly how the inversion may be accomplished, 
make the substitutions
\begin{eqnarray}
k_{vx} &=& - s \cos 2\pi \phi, \\
k_{vy} &=& + s \sin 2\pi \phi, 
\end{eqnarray}
on the right-hand side of Eq.~\ref{eq:2dFT}. This
leads to a factor $2\pi |s|$ from the Jacobian of the coordinate transform, and allows the Fourier transform of the image to be 
expressed in terms of the line profiles using Eq.~\ref{eq:2dkey}.
One finds
\begin{equation}
I_2(v_x, v_y) = 2\pi \int_0^\infty ds \int_0^1 d\phi \, 
|s| \tilde{f}(s,\phi) e^{i 2\pi s V_R}, \label{eq:2dinvert}
\end{equation}
where $V_R$ is a function of $\phi$ as given by Eq.~\ref{eq:straight}. This can be re-written as
\begin{equation}
I_2(v_x, v_y) = 2\pi \int_0^{0.5} d\phi \int_{-\infty}^{\infty} ds \,
|s| \tilde{f}(s,\phi) e^{i 2\pi s V_R},
\end{equation}
which can be broken down into two steps: first, an inverse Fourier transform  step that returns a filtered version of the line profiles
\begin{equation}
F(V,\phi) = \int_{-\infty}^{\infty} ds \,
|s| \tilde{f}(s,\phi) e^{i 2\pi s V},
\end{equation}
and, second, an integral over orbital phase
\begin{equation}
I_2(v_x,v_y) = 2\pi \int_0^{0.5} d\phi \, F(-v_x \cos 2\pi\phi + v_y\sin2\pi \phi, \phi) .
\end{equation}
This last step is known as "back-projection", because the contribution to the image of a particular phase can be imagined as smearing the line profile back over the image in the same direction as the projection that formed the profile in the first place (see \citet{2001LNP...573....1M} for a pictorial representation of the process).

Subject to issues such as full phase coverage, spectral and temporal resolution and signal-to-noise, this is a well-defined procedure with relatively little room for a multiplicity of different images that can match a given set of line profiles.
The key result leading to this is Eq.~\ref{eq:2dkey} which allows the substitution of the Fourier transform of the image in terms of the line profile in Eq.~\ref{eq:2dinvert}. The 3D case is a simple extension of the procedure leading up to Eq.~\ref{eq:2dkey}, and it leads to a very similar looking result, but in this case the full 3D transform is not obtained, and this makes a critical difference to the computation of 3D maps.

\section{Line formation in 3D velocity space}
\label{sec:3dlform}

The extension to 3D follows very much the same lines as the previous section.
The same quantities appear, but without the subscripts "2" -- unadorned quantities
are three dimensional. Points in 3D velocity are labelled by 
$\vvec = (v_x,v_y,v_z)$. The $x$- and $y$-axes are defined as before, while the additional $z$-axis is defined to complete a right-handed triad, i.e. $\hat{\mathbf{z}} = \hat{\mathbf{x}} \times \hat{\mathbf{y}}$, where carets denote unit vectors. Thus defined, the $z$-axis is parallel to the angular momentum vector of the binary, and therefore the orbital inclination $i$ is the angle between the $z$ axis and the vector
$\hat{\mathbf{e}}$ pointing from the binary towards Earth, i.e.
$\cos(i) = \hat{\mathbf{z}} \cdot \hat{\mathbf{e}}$.
Thus as well as the restriction $i\neq\ang{0}$ noted earlier, we must also have $i\neq \ang{90}$,
as has been remarked previously \citep{2006ApJ...652.1547A}.

The line profile $f(V,\phi)$, a function of radial velocity $V$ and orbital phase $\phi$, due to image $I$ in 3D velocity space is given by
\begin{eqnarray}
f(V,\phi) &=& 
\iiint dv_x\,dv_y\,dv_z\, \delta(V-V_R) I(v_x,v_y,v_z), \\
&=&\int_\vvec d^3\vvec\, \delta(V-V_R) I(\vvec), 
\label{eq:pform}
\end{eqnarray}
adopting the same compact notation as before. The radial velocity $V_R$ 
is now given by
\begin{eqnarray}
V_R &=& - v_x \cos (2\pi \phi) + v_y \sin (2\pi \phi) - v_z , \label{eq:3dv}\\
    &=& \uvec \cdot \vvec, 
\end{eqnarray}
where the vector $\mathbf{u}$ is defined as
\begin{equation}
    \uvec = (-\cos (2\pi \phi), \sin(2\pi \phi), -1).
\end{equation}
As for the 2D case, $v_x$ and $v_y$ implicitly include a $\sin(i)$
projection factor along the line-of-sight, but now the $v_z$ component implicitly includes a $\cos(i)$ projection factor for the reasons outlined before.

Like Eq.~\ref{eq:p2form}, Eq.~\ref{eq:pform} expresses a form of projection, but now from three dimensions to one dimension. The delta function again selects all points in the (now 3D) image that have radial velocity $V_R = V$ at orbital phase $\phi$. For a given orbital phase, the line profile can be thought of as the result of slicing the 3D image into a series of flat, parallel slices and integrating over the two dimensions spanning each slice to end up with a one dimensional function for a given $\phi$. The orientation of the planes is governed by their perpendicular vector, \uvec. Vector \uvec\ has unit length when projected into the $(v_x,v_y)$ plane, and a $v_z$ component of the same length. Therefore it makes an angle of $45^\circ$ with 
the $v_z$ axis and also with the $(v_x,v_y)$ plane. Imagining it as an arrow starting from the origin in velocity space, \uvec\ sweeps out a cone with an opening angle of \ang{45} around the $v_z$ axis as the binary rotates. This geometry feeds directly through to the reciprocal Fourier space as will soon be shown, and shows up later in the reconstructions.

\begin{figure}
	\includegraphics[width=\columnwidth]{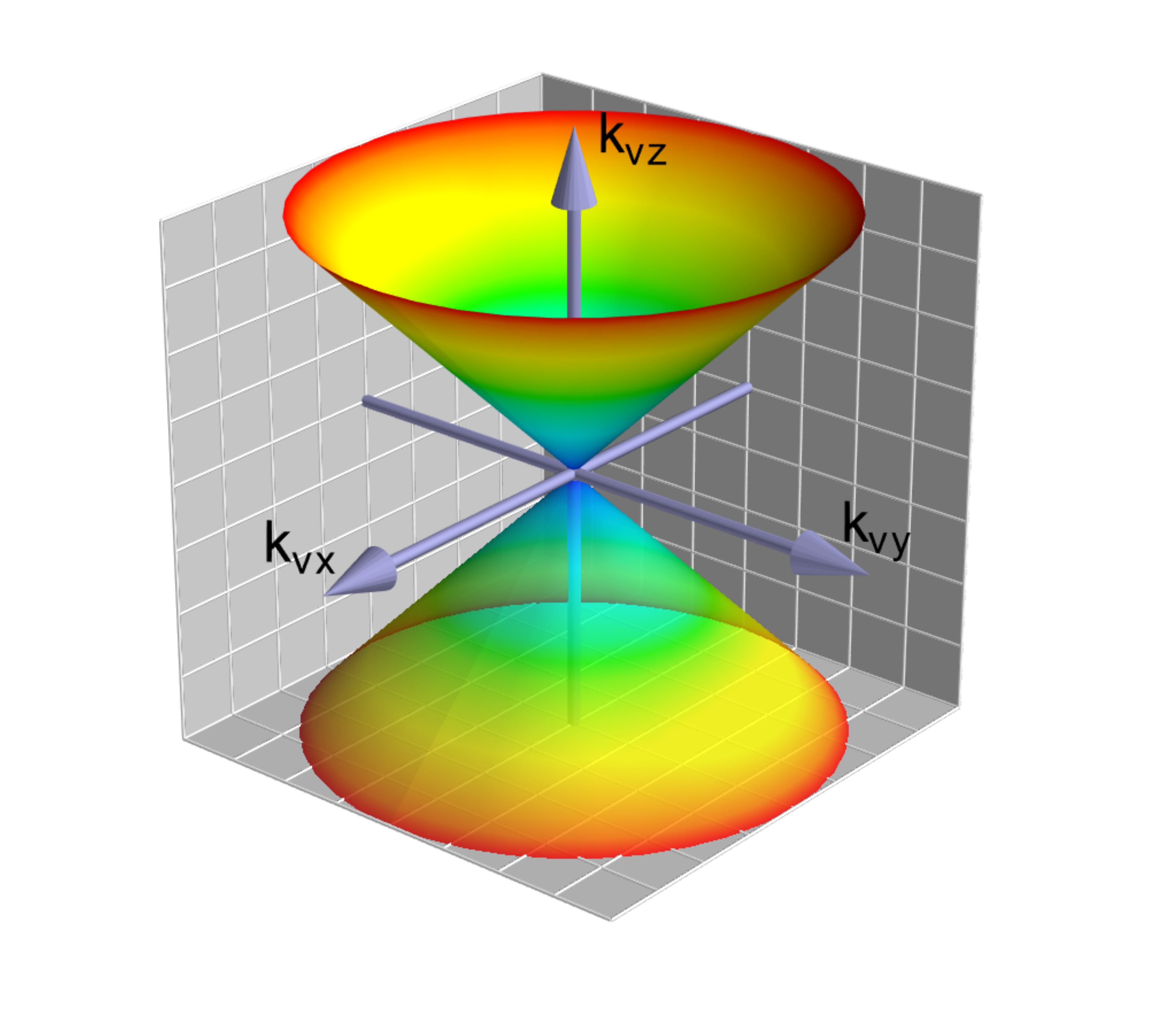}
    \caption{An illustration of the double cone surface in Fourier \kvec-space over which the values of the Fourier transform of a 2D velocity-space Doppler image are constrained by atomic line profiles at all phases of a binary star. The double cone has its apex at the origin $\kvec = (0,0,0)$, an  opening angle of $45^\circ$, and it is aligned with $k_{vz}$ axis. (The figure was created using the 3D visualisation software, \protect{\tt mayavi}, \citet{2011CSE....13b..40R}.)
    \label{fig:dcone}}
\end{figure}

Moving to Fourier space as before, but now in 3D,
\begin{eqnarray}
\tilde{I}(\kvec) &=& \int_\vvec d^3\vvec \, e^{- i 2\pi \kvec \cdot \vvec} I(\vvec),\\
I(\vvec) &=& \int_\kvec d^3\kvec \, e^{i2\pi \mathbf{k}\cdot\mathbf{v}} \tilde{I}(\mathbf{k}), \label{eq:3dFT}
\end{eqnarray}
where $\tilde{I}(\mathbf{k})$ is the 3D Fourier transform of the 3D velocity space image.  Substituting for $I(\vvec)$ using Eq.~\ref{eq:3dFT} leads as before to
\begin{equation}
f(V,\phi) = \int_\kvec d^3\kvec \int_\vvec d^3\vvec \, \delta(V-\uvec\cdot\vvec) e^{i2\pi \kvec\cdot\vvec} \tilde{I}(\kvec), 
\end{equation}
and then the same substitution of the delta function and subsequent manipulation as applied to Eq.~\ref{eq:2dinter} again leads to
\begin{equation}
    \tilde{I}(s {\mathbf u}) = \tilde{f}(s, \phi).
    \label{eq:key_result}
\end{equation}
This is identical in form to its 2D equivalent, Eq.~\ref{eq:2dkey}, apart from the loss of the subscripts on $I$ and \uvec. However, because of the switch from 2D to 3D, its implications for inversion are very different. As before, the interpretation of the relation is that the Fourier transform with respect to radial velocity of the line profile at phase $\phi$ gives us the values of the Fourier transform of the
image that we are after, along lines through the origin in \kvec-space of direction \uvec. The difference is that, in 2D, as the binary orbits, these lines cover the whole of 2D \kveca-space. In the 3D case by contrast, the lines span the surface of a double cone of opening angle \ang{45}, with its apex at the origin of \kvec-space and with the $k_{vz}$ axis as the cone's axis. This is illustrated in Fig.~\ref{fig:dcone}. 
In the 3D case we can only measure the values of the Fourier transform on a 2D surface embedded within 3D \kvec-space.

Since we do not recover the 3D Fourier transform throughout \kvec-space, we do
not have the information required to deduce $I(\vvec)$ from a 3D
inverse Fourier transform, and therefore no unique inversion along the
lines of 2D Doppler tomography is possible. Hence 3D tomography should not be viewed as a more advanced form of 2D tomography -- the two cases are fundamentally different in nature.

This result means that one
can generate an infinite number of very different 3D images that are observationally indistinguishable because they each produce identical
line profiles. Consider for
instance the following procedure: (i) select an image $I(\vvec)$; (ii)
compute its Fourier transform $\tilde{I}(\kvec)$; (iii) add a modifier function, $\tilde{M}(\kvec)$ to $\tilde{I}(\kvec)$ subject to the
restriction that it is zero on the surface of the double cone discussed before; (iv) invert the modified Fourier transform to generate a new image, $I'(\vvec)$. This procedure guarantees that the line profiles corresponding to $I$ and $I'$ will be identical.

In addition to the restriction that it is zero on the
double cone, the function $\tilde{M}$ of step (iii) is subject to
well-known symmetry requirements to keep its inverse Fourier transform
real, e.g.
$\tilde{M}^*(k_{vx},k_{vy},k_{vz})=\tilde{M}(-k_{vx},k_{vy},k_{vz})$,
where the asterisk denotes the complex conjugate; similar conditions apply to the other two components.

It is not hard to think of functions that satisfy these conditions; an explicit example will be shown in section~\ref{sec:spot} later. There is however an important additional constraint that may come to our aid: positivity. As well as being real (i.e. $I = I^*$), physical emissivity distributions should
satisfy the condition $I(\vvec)\ge 0$ for all $\vvec$. If the initial
image chosen in step (i) has large regions of zero or near zero flux,
then the possibilities for the function $\tilde{M}$ become much more
restricted. We may well expect that in many instances most of the
volume of 3D Doppler images is indeed rather empty, so this is a
significant point in our favour. It does suggest however that the
potential for development of artefacts in 3D images will depend upon
the nature of the image itself, and this is something that will be
seen in section~\ref{sec:simulated}, where the artificial reconstructions are presented.

\subsection{A return to 2D}

The 2D case can be recovered from the work of the preceding section if
we set the 3D image to be of the form
\begin{equation}
    I(\vvec) = I_2(v_x, v_y) \, \delta(v_z),
\end{equation}
i.e. we allow no motion out of the orbital plane (and for simplicity assume zero systemic velocity, with no loss of generality). With these assumptions, the Fourier transform becomes independent of $k_{vz}$
\begin{equation}
    \tilde{I}(\kvec) = \tilde{I}_2(k_{vx},k_{vy}),
\end{equation}
and Eq.~\ref{eq:key_result} becomes identical to its 2D equivalent, Eq.~\ref{eq:2dkey}. With no motion out of the plane allowed, the values of the Fourier transform at the two points on the double cone for a given pair of  $(k_{vx},k_{vy})$ values are the same, and they are the same as the value at $(k_{vx},k_{vy},0)$, hence we know the 2D transform in the plane $k_{vz} = 0$, which gets us back to the well-constrained 2D case.

\section{Can we say anything about the 3D emissivity distribution?}
\label{sec:anything}

Although it has just been shown that an inversion along the lines of 2D tomography is not possible, it does not mean that it is not possible to find \emph{an} emissivity distribution in 3D corresponding to a given set of line profiles. In the context of accreting binary stars, the situation is somewhat analogous to the use of light curves in eclipse mapping \citep{1985MNRAS.213..129H}. The information provided by light curves during ingress and egress is equivalent to 2D-to-1D projections of the accretion disc at just two angles. This corresponds to knowing the values of the 2D Fourier transform along just two lines out of the entire plane of possible values, and yet eclipse mapping has proved useful in understanding the emissivity distributions of discs. Thus all is not necessarily lost as a result of Eq.~\ref{eq:key_result}, even though it is clear that the inversion of the 3D case will have a different character from its 2D counterpart, and a key question becomes whether one can deduce anything of use about the 3D distribution from line profile data.

\begin{figure*}
	\includegraphics[width=0.9\textwidth]{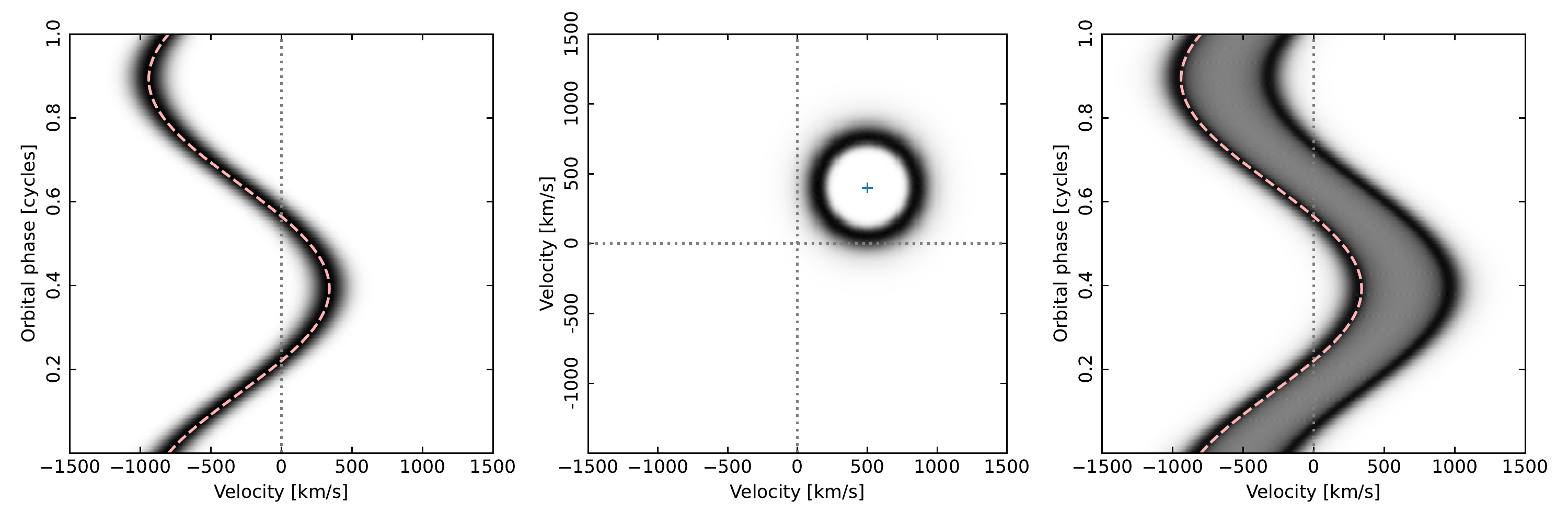}
    \caption{\emph{Left:} lines profiles plotted in the form of a trailed spectrum of orbital phase versus radial velocity. They correspond to a gaussian blob of emission with Full Width at Half Maximum (FWHM) of \SI{300}{\km\per\s} centred on $(V_x,V_y,V_z) = (500,400,300)\,\si{\km\per\s}$. The dashed line shows the exact path expected of the centre of the distribution.
    \emph{Centre:} the image that results from applying standard 2D Doppler tomography to the profiles of the left panel takes the form of a ring of radius $V_Z = \SI{300}{\km\per\s}$ centred upon $(V_X,V_Y) = (500,400)\si{\km\per\s}$ (marked by the plus sign). In 3D, the ring is the cross-section of a cone of opening angle $45^\circ$ degree with apex at the centre of the gaussian blob in 3D and axis parallel to $v_z$ with the $v_x$--$v_y$ plane.  \emph{Right:} The profiles computed from the 2D image are a very poor fit to the data. Dotted lines are used here and in other figures to mark zero velocity in the images and zero radial velocity in the data.
    \label{fig:single}}
\end{figure*}

Although far from complete, it is clear that \emph{some} information on the $v_z$ component is encoded in line profiles. Consider for instance a 3D gaussian blob of emission centred at $\vvec = (V_x,V_y,V_z)$ with $V_z \neq 0$. This will appear in spectra as a gaussian emission-line profile with a mean offset of $-V_Z$ from zero velocity, which varies sinusoidally in velocity around the mean with amplitude $\sqrt{V_x^2+V_y^2}$. The left panel of Fig.~\ref{fig:single}
shows an example of line profiles from such an image, plotted in the traditional form of a trailed spectrum with orbital phase running up the page. The standard 2D inversion, using the maximum entropy method with positivity enforced \citep{1988MNRAS.235..269M}, is shown in the centre panel of the figure. A volcano-like ring is formed as the result of the substantial offset in the $v_z$ direction, a structure discussed by \citet{1988MNRAS.235..269M} in the context of the effect of incorrect systemic velocities upon 2D maps. The ring allows the appearance of an offset sinusoidal component matching the data, but at the expense of a second sinusoid and intervening emission, neither of which are present in the data of the left-hand panel. The key point is that the data corresponding to the image (right-hand panel) are an extremely bad fit to the input data. Thus a 2D map cannot in general simply adjust itself to fit line profiles that originate from an intrinsically 3D distribution, and the 3D nature of the input map is not entirely lost in the process of line profile formation.

It is however also evident that there are cases where, without further prior constraints, more than one very different distribution in 3D can lead to identical sets of profiles, i.e. there is a fundamental degeneracy that reflects the loss of information represented by Eq.~\ref{eq:key_result}. This was discussed in terms of 
applying modifications in Fourier space earlier, but there is a simple explicit instance of degenerate images which should reinforce the point.

Consider a 2D distribution, i.e. the traditional situation with no motion out of the plane, but restrict it to exact axi-symmetry about the origin. Such a distribution projects to a line profile that is independent of orbital phase given by
\begin{equation}
    f(V) = \int_{-\infty}^{\infty} du \, I_2\left(\sqrt{V^2+u^2}\right),
\end{equation}
where now the 2D emissivity depends only upon the distance from the origin. This profile is even about $V=0$, but otherwise can take a wide variety of forms depending upon $I_2(V)$. However, no matter what form $f(V)$ takes, it can also be generated from an entirely different distribution of the form
\begin{equation}
    I(\vvec) = f(v_z) \, \delta(v_x)\,\delta(v_y),
\end{equation}
because substituting this into Eq.~\ref{eq:pform} and, with the help of Eq.~\ref{eq:3dv}, carrying out the integrals first over $v_x$ and $v_y$, and finally over $v_z$, leads to
\begin{eqnarray}
f(V,\phi) &=&\int_\vvec d^3\vvec \, \delta(V-V_R) \delta(v_x)\, \delta(v_y),\\
&=& \int_{-\infty}^\infty dv_z\, f(v_z) \delta(V+v_z),\\
&=& f(-V) = f(V),
\end{eqnarray}
using the even nature of the profile in the last line. 

Hence in this restricted case, the same set of line profiles, represented by the even function $f(V)$, independent of phase, can be explained by either a 2D distribution confined to the $v_x$--$v_y$ plane or equally by a 1D distribution along the $v_z$ axis. This is familiar from real systems where it is not always clear whether one is looking at line profiles from a flat accretion disc or from symmetric jets along the orbital axis. In practice, the form of emission, disc or jet, is usually evident for other reasons, but treated purely in terms of inversion into 3D, the degeneracy is clear.

In summary, line profiles carry some information about the emissivity distribution in 3D, but at the same time there is scope for degeneracy and hence for false structures to be generated during inversion. The significance of the degeneracy depends upon the exact distribution of emission in 3D.
Bearing this very real pitfall in mind, I now look at a practical implementation of 3D inversion.

\section{Implementation of 3D inversion}
\label{sec:implement}

The 3D inversion adopted in this paper is a direct extension of the maximum entropy method presented by \citet{1988MNRAS.235..269M}. The key to that 
method is a routine which calculates data (i.e. the line profiles) corresponding to an image, an operation which can be expressed as a matrix operation, i.e.
\begin{equation}
    d_i = A_{ij} I_j, \label{eq:matrix}
\end{equation}
where $I_j$, $j = 1$ to $N$, are the $N$ elements of the image and $d_i$, $i = 1$ to $M$, are the $M$ data points, and the matrix $\mathbf{A}$ encapsulates the profile formation by projection described earlier. Summation over the index $j$ is assumed. The MEMSYS algorithm \citep{1984MNRAS.211..111S} adopted to implement Doppler tomography requires a function to implement Eq.~\ref{eq:matrix} and a closely-aligned function, with very similar looking code, to effect its transpose.

In 2D tomography the image is represented by a square, 2D array of typically 100x100 to 400x400 pixels. Thus the number of elements $N$ typically ranges from 10,000 to 160,000. 3D tomography merely requires the image to become three dimensional. There is no need for the $v_z$ dimension to match the $v_x$ and $v_y$ dimension, and thus one may have 300x300x100 elements for instance. One way to think of the 3D image is as a series of 2D image slices displaced in terms of their systemic velocity, and this is effectively how the calculations are implemented in practice. The computational time and storage requirements increase in proportion to the $v_z$ dimension, but computer speeds and memory capacity have advanced enormously in the decades since \citet{1988MNRAS.235..269M}, and the process can be effectively parallelised across multiple cores, so, although it can take several seconds to compute the data equivalent to a 3D image, the greatly increased computational burden is not a fundamental barrier. The code is wrapped in a {\tt Python} package and available via {\tt github}\footnote{https://github.com/trmrsh/trm-doppler}. Further information on its features, which include several additional advantages over \cite{1988MNRAS.235..269M}'s original implementation, is left to appendix~\ref{app:dcode}.

An aspect of the maximum entropy method which is of somewhat secondary importance in the 2D imaging case, the "default image", is of much greater significance in the 3D case. The default image $J$ enters into the computation of entropy
\begin{equation}
    S = -\sum_{i=1}^N p_i \log \left( \frac{p_i}{q_i} \right), \; \text{where} \;
    p_i = \frac{I_i}{\sum_j I_j},\; \text{and} \; q_i = \frac{J_i}{\sum_j J_j},
\end{equation}
and allows one to control the features of the image that one wants to have the largest weight when computing the entropy. In the absence of any data constraints, the image of maximum entropy is the default image, i.e. $I = J$, so $S$ is a measure of how far the image deviates from the default, and the idea is to build into the default image aspects one expects the image to show.

In 2D Doppler, imaging the standard default $J$ is a gaussian blurred version of $I$. This is isotropic and does not favour any particular structure or direction within the image, and it makes the entropy primarily sensitive to short-scale noise in the image, with larger scale structure entirely determined by the data. This is useful given the strongly-constrained nature of the 2D case. In the case of disc eclipse imaging by contrast, a more constraining azimuthal default, computed from the radial profile of the image, is often applied \citep{1985MNRAS.213..129H}. This partly reflects the relatively incomplete constraints set by the data in this case, as discussed earlier. During optimisation, the default is taken to be constant by the MEMSYS3 code used, thus the default always needs to be re-computed and the optimisation re-run until the point is reached at which no significant changes take place. This procedure was adopted for all the computations shown below. The changing nature of the default is one of the main drivers of computation time as it can sometimes take many iterations to achieve a near steady state, although it has to be said that very often there is no noticeable change in the visual appearance of the reconstructed images starting well before such a state is reached.

\section{Recovery of simulated images}
\label{sec:simulated}

In this section artificial images are used to generate data from which recovery of the input image is attempted. A 3D version of the gaussian blurring was the first method adopted for default computation. The blurring used for this was accomplished using a Full Width Half Maximum (FWHM) of \SI{200}{\km\per\s} along all axes. An alternative procedure will be introduced in section~\ref{sec:disc}, when disc imaging is presented. The 3D images shown next had dimensions of $(400,400,300)$ (i.e. 300 slices in the $v_z$ direction), with voxels spaced by \SI{10}{\km\per\s} along all 3 axes.

The procedure followed was first to create an artificial model image, and then generate line profile data from it, with the addition of a small amount of pseudo-random noise. The line profiles were computed at 200 phases, equally spaced around an orbit, and placed on a wavelength scale with 1000 pixels, each  \SI{6}{\km\per\s} in width, with an assumed instrumental resolution of FWHM $=\SI{20}{\km\per\sec}$. The inversions were based upon these data, starting with initially uniform images. 

In all cases, it is good to keep in mind the bi-conical structure discussed for the profile formation from 3D images, because it is this which largely determines the structure of artefacts in the reconstructions. One way to look at this is to imagine that there is a tendency for any given feature on an image to spread out into a double cone extending up and down from it in the $v_z$ direction. This will be seen in some of the projections taken perpendicular to the $v_z$ axis, and will be seen to particular effect in the case of disc emission. Just this was seen already in the centre panel of Fig.~\ref{fig:single}, which is a cross section perpendicular to the axis of the double cone spooled by the $v_z \neq 0$ model spot used to create the data.

The images chosen were selected in large part for simplicity. 3D images are best appreciated through "live" dynamic movie-style rendering and when fixed into 2D form, they can become difficult to interpret. Thus I have adopted a simple approach here of showing them in projection (i.e. summed along the suppressed axis) or occasionally
in slices because this makes direct side-by-side comparison much easier. As a consequence of this I have selected simple shapes where it is not hard to see how they should appear in such plots. Some of them approximately conform to structures one can expect to see in real systems, but I have not attempted to simulate such structures closely as then it becomes hard to unravel the physics from the issues of inversion, with the latter being the chief concern of this paper.

\subsection{Gaussian spot}
\label{sec:spot}
\begin{figure}
	\includegraphics[width=\columnwidth]{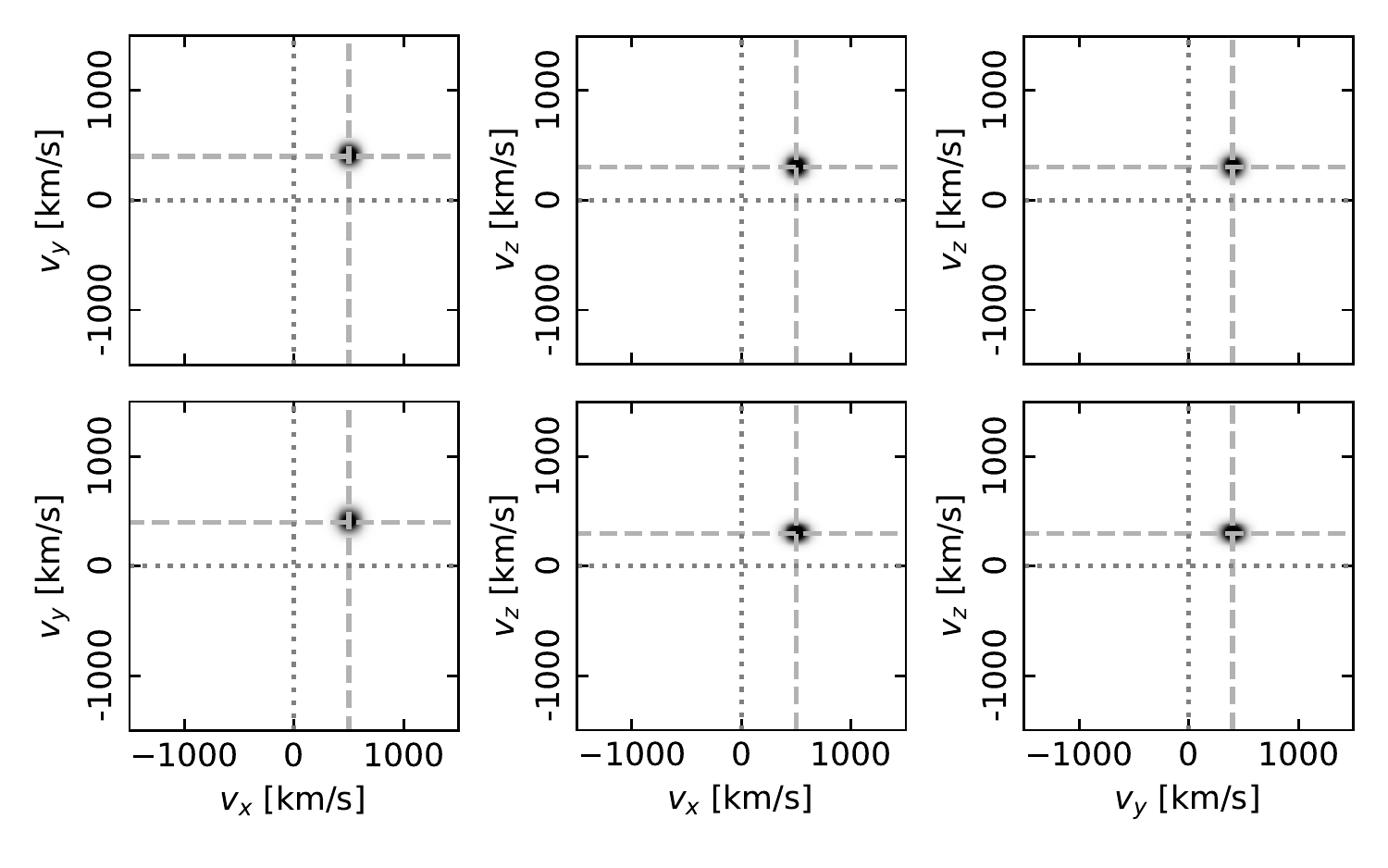}
    \caption{\emph{Top row:} projections of the 3D model image (a gaussian spot) used to generate the data of Fig.~\protect\ref{fig:single}. From left to right, the projections show the $v_x$--$v_y$ plane, the $v_x$--$v_z$ plane and the $v_y$--$v_z$ plane. \emph{Bottom row:} the same projections of the 3D reconstructed image. The dashed lines mark the central velocity of the model spot.
    \label{fig:single-proj}}
\end{figure}

The first test was the attempted recovery of the gaussian blob used to generate the data shown in the left panel of Fig.~\ref{fig:single}, data for which 2D tomography comes up short. 
The 3D reconstruction is compared against the original image in the form
of projections along the $v_z$, $v_y$ and $v_x$ axes in Fig.~\ref{fig:single-proj}.
I don't show the reconstructed data in the 3D case, since they are visually indistinguishable from the input data shown on the left of Fig.~\ref{fig:single}.
Given what has been said about the impossibility of a full inversion in the 3D case, the image appears to have been recovered remarkably well, although the reconstructed images are more extended in the $v_x$-$v_y$ plane than in the $v_z$ direction compared to the model. The images shown are projections in which the 3D images have been summed along the missing axis for each image, which can potentially be misleading as a comparison, but in this case individual slices also compare very well, and at the velocity of peak flux look very similar to Fig.~\ref{fig:single-proj}.

\begin{figure}
	\includegraphics[width=\columnwidth]{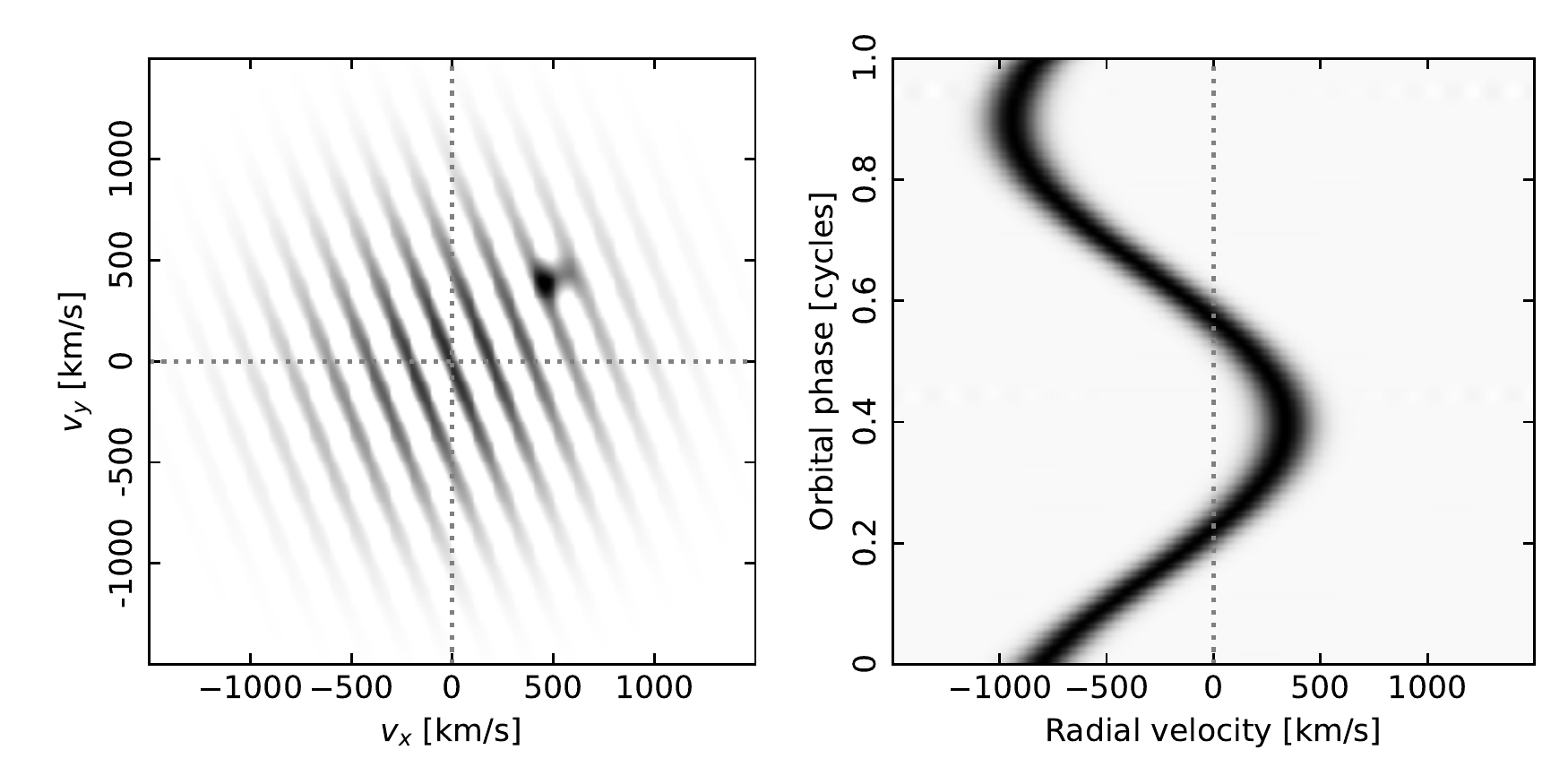}
    \caption{\emph{Left:} a $v_x$--$v_y$ slice through the gaussian spot at $v_z = \SI{300}{\km\per\sec}$ where it is strongest, after addition of a tapered plane wave as discussed in the main text. \emph{Right:} the corresponding line profiles, which should be compared to the line profiles shown in the left panel of Fig.~\protect\ref{fig:single}. The figure is an explicit example of the multiplicity of 3D images that can match the same set of data.    \label{fig:single-mod}}
\end{figure}

Before moving on, the elementary spot is a good chance for an explicit demonstration of the fact that very different images in 3D can result in the same line profiles. To show this, I added a harmonic plane wave to the elementary 3D spot image with a phase $\psi = 2\pi \kvec \cdot \vvec$ (in radians), where I chose
\begin{equation}
    \kvec =(0.005,0.002,0.001)\; \si{\per\km\s},
\end{equation}
The Fourier transform of such a wave consists of two delta functions located at $\pm \kvec$. The coefficients of $\kvec$ above were picked to ensure that these points did not lie close to the double cone in \kvec-space, i.e. such that 
\begin{equation}
    k^2_{vx} + k^2_{vy} \neq k^2_{vz}.
\end{equation}
In addition I applied a gaussian taper to the amplitude of the plane wave of the form $\exp(-(v/500)^2/2)$. This substantially reduces sharp edge ringing effects, since for practical reasons, the images span only a finite range of velocity space. Such a taper corresponds to convolution by a gaussian in \kvec-space, and, as long as it does not spread as far as the double cone, ensures the condition that the values of the Fourier transform on the cone are not altered. Once the image had been altered in this manner, I computed line profiles from it. The results of this are shown in 
Fig.~\ref{fig:single-mod}.
As it was designed to do, the addition of the plane wave to the image has made no visible difference to the line profiles which match those of the unadulterated model shown in the left panel of Fig.~\ref{fig:single}. This is very different from the 2D case.  Were the single slice shown in Fig.~\ref{fig:single-mod} a 2D image, the projections at phases aligned with the wave peaks and troughs would evidently show a high amplitude sinusoid. It is striking how the 3D image can be modified with a strong artefact that almost obscures the "real" spot feature, even at its strongest, and yet in the data, only the spot can be seen. Of course, the plane wave violates positivity in this case, but it is not clear that it always would do so if other patterns were considered.

\subsection{Uniformly-filled cube}

\begin{figure}
	\includegraphics[width=\columnwidth]{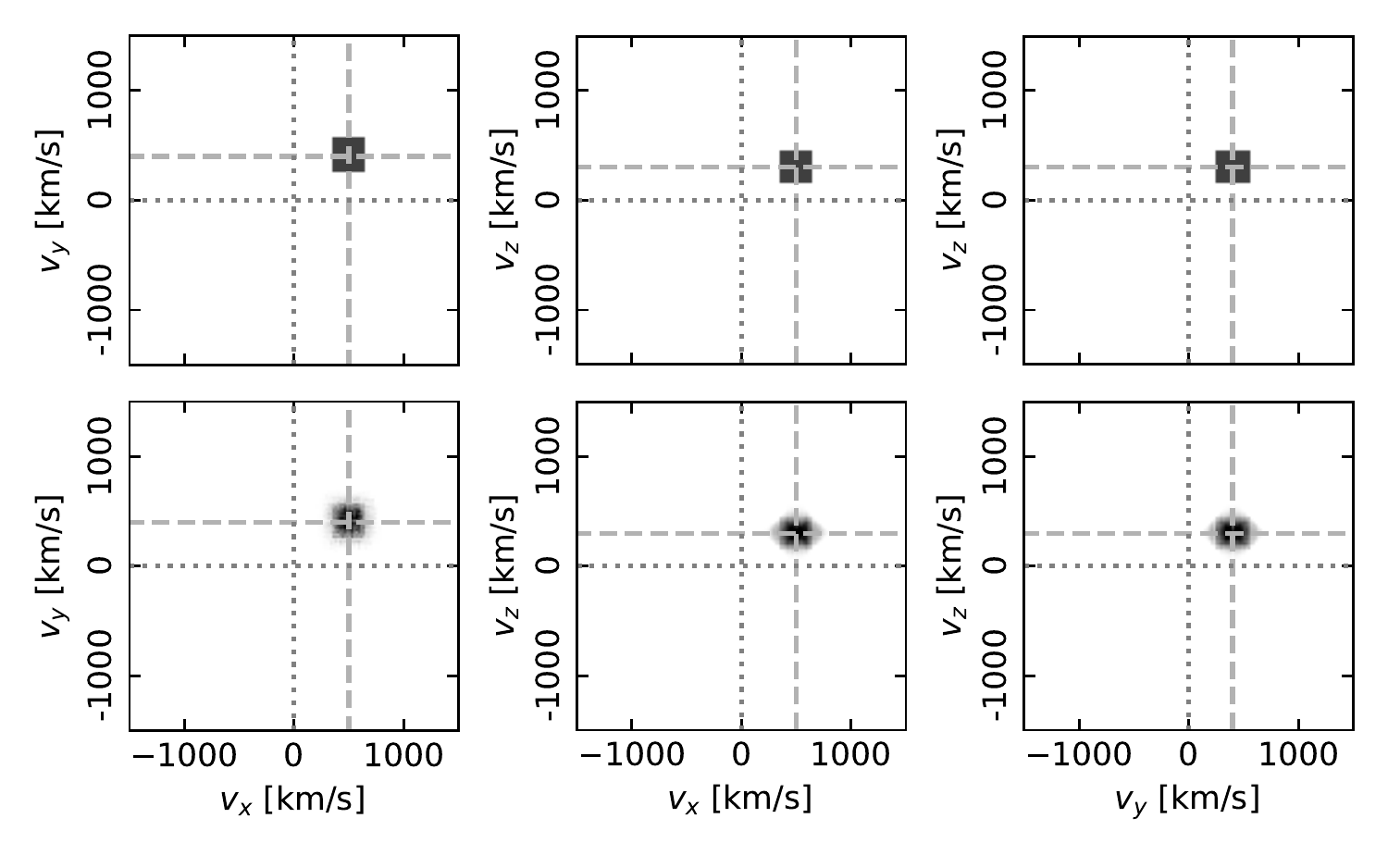}
    \caption{\emph{Top row:} projections of the 3D model image (a uniformly filled cube). From left to right, the projections show the $v_x$--$v_y$ plane, the $v_x$--$v_z$ plane and the $v_y$--$v_z$ plane. \emph{Bottom row:} the same projections of the 3D image reconstructed from data computed from the model image. The dashed lines mark the central velocity of the model spot.
    \label{fig:cube-proj}}
\end{figure}

The gaussian spot perhaps provides a soft start for the 3D reconstruction given the use of a gaussian convolution to derive the moving default images. To provide a tougher test, a test image was constructed with a spot in the form of a cube of dimension \SI{300}{\km\per\sec} on all sides centred on the same location as used for the gaussian spot, i.e. $(500,400,300)\,\si{km\per\s}$. As before, the reconstruction employed gaussian convolution. The result, again shown with three projections as in Fig.~\ref{fig:single-proj}, is shown in Fig.~\ref{fig:cube-proj}.
In this rather high signal-to-noise simulation it proved difficult to achieve a very good fit to the data as the optimisation became very slow to reduce the $\chi^2$ in the later stages. This is at least partly a consequence of the unusual nature of the input image which features significant step changes in intensity, very much against the spirit of a gaussian-convolved default, and one can guess that the direction of maximum entropy is almost anti-parallel to the direction of decreasing $\chi^2$. It is probably also indicative of the nature of the lesser constraints in 3D compared to 2D. Nonetheless,
the cubic shape of the model image certainly shows through in the reconstruction, and although it is noticeably imperfect, it seems to be a reasonably useful reflection of the input model.

\begin{figure}
	\includegraphics[width=\columnwidth]{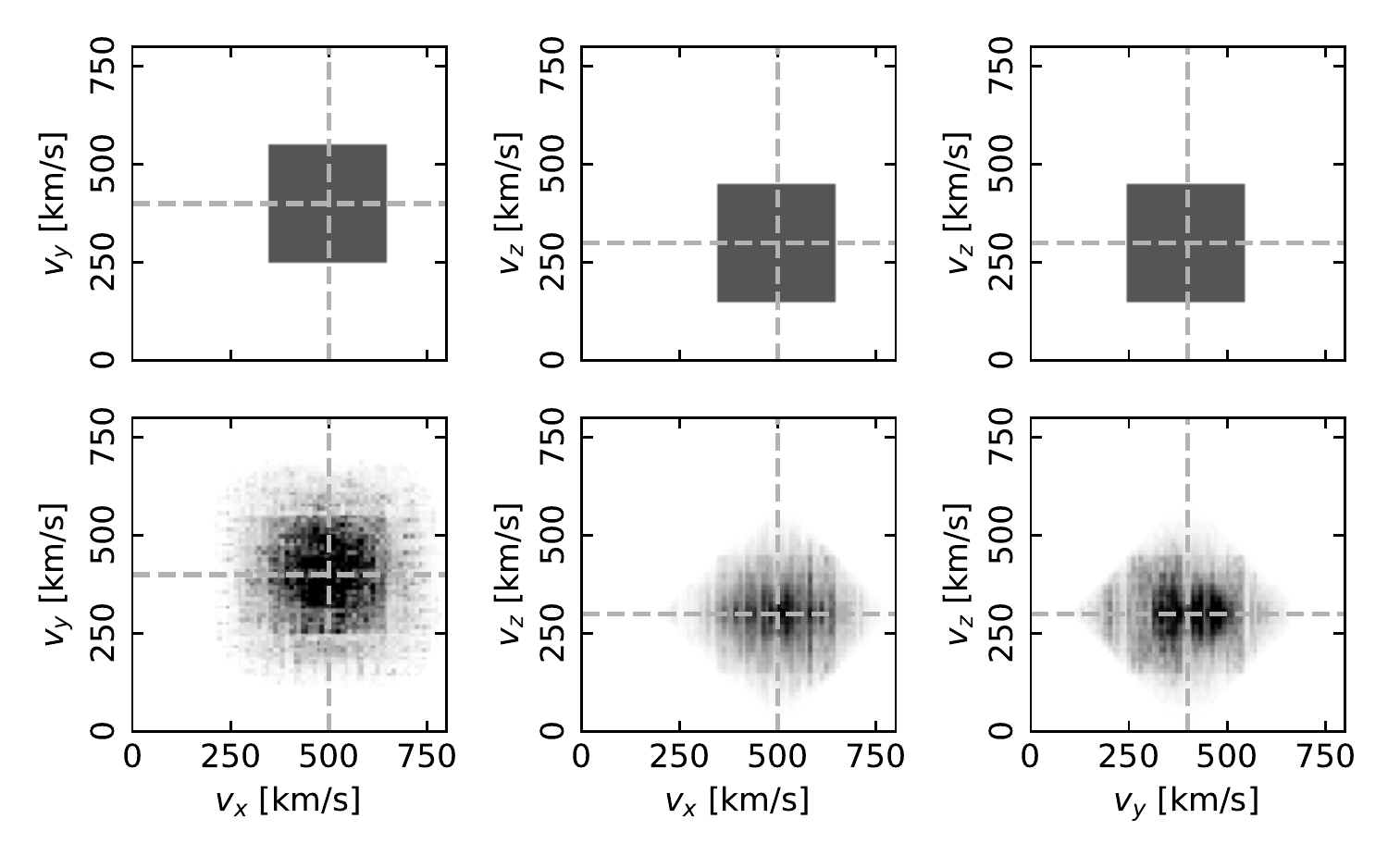}
    \caption{A zoomed-in display of the centre-most 2D slice of the 3D model image (top) and the reconstruction (bottom) for the uniformly filled cube simulation. The same intensity levels
    have been used in each case. \label{fig:cube-slice}}
\end{figure}

In this case the use of projections does mask some problems. Fig.~\ref{fig:cube-slice}
shows just the individual slices closest to the centres of the cubes in their respective directions.
This shows very fine structure, much of which is generated as the result of the cliff-edge nature of the image, but in addition this is the first clear example of the bi-conical structure mentioned at the start of the section. Looking at the two panels in the bottom right of the figure, the original square cross-section of the cube can be dimly glimpsed, but somewhat more obvious is an outer square at \ang{45} to the model square, and encompassing it. This is a consequence of the bi-conical smearing mentioned before. The same effect in the $v_x$--$v_y$ plane blurs the outline of the square as well, but in a different manner owing to the anisotropic nature of the double cone. 

Despite the evident artefacts, one could in this case correctly deduce the existence of emission at the location of the cube, even if its details would be unreliable.

\subsection{A 2D disc imaged in 3D}
\label{sec:disc}

\begin{figure}
	\includegraphics[width=\columnwidth]{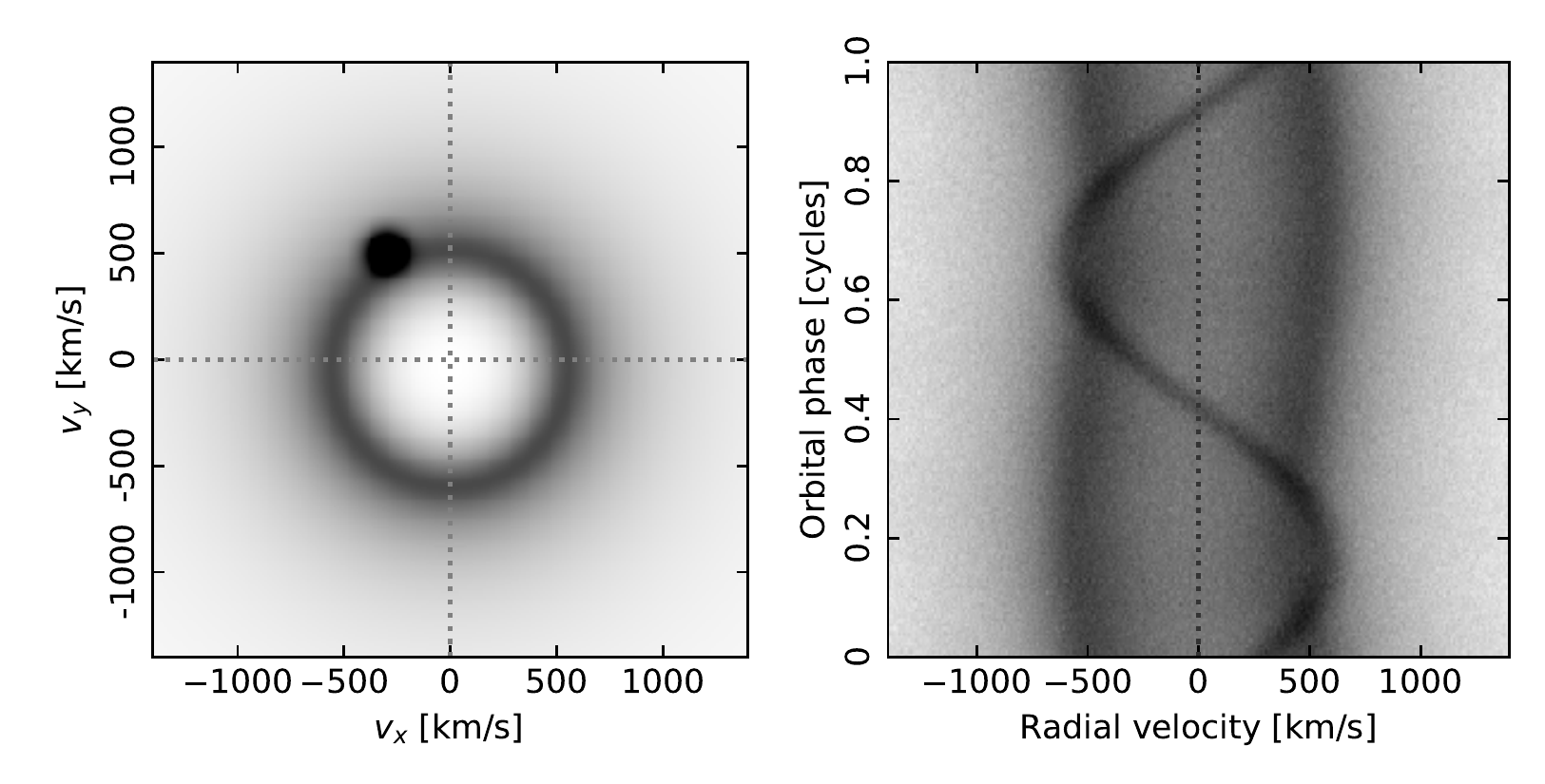}
    \caption{An artificial accretion disc plus bright-spot (left) and data computed from them (right). The disc is centred at $(0,-50)\,\si{\km\per\s}$; the spot is centred at $(-300,500)\,\si{\km\per\s}$.
    \label{fig:disc_model_data}}
\end{figure}

\begin{figure}
	\includegraphics[width=\columnwidth]{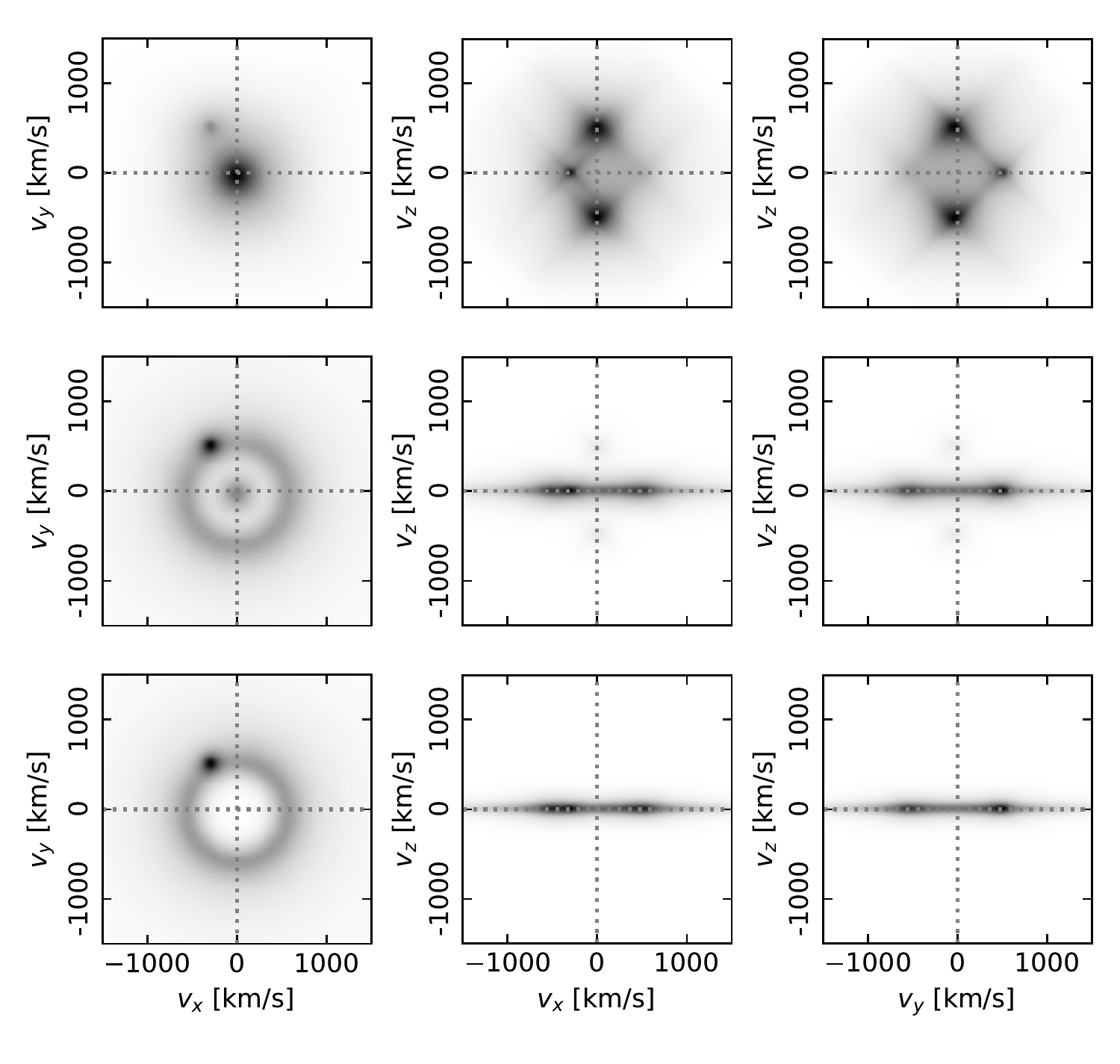}
    \caption{3D reconstructions created from the data generated from the 2D map and bright-spot of Fig.~\protect\ref{fig:disc_model_data}. They are shown as projections. In each case the same $\chi^2$ per data point ($=1.00$) was achieved. \emph{Top row:} reconstruction using a standard 3D gaussian convolution.
    \emph{Centre row:} reconstruction with a 25\% "pull" towards a 2D image (see main text). \emph{Bottom row:} reconstruction with a 50\% default pull.
    \label{fig:disc-proj}}
\end{figure}

The previous two simulations have involved images both of which had most of their flux concentrated into one small region. Reconstruction in such cases has the most to gain from the positivity condition discussed in section~\ref{sec:3dlform}, a condition that is explicitly built into the simplest form of the maximum entropy inversion. Positivity suppresses any artefacts which would cause parts of the image to be negative. Therefore, to move a little away from that, and to treat a case that will undoubtedly come up in practice, I now look at emission from a disc. A standard 2D image of a disc plus a bright spot was created and data computed from it in the manner specified earlier. The model image and corresponding data are shown in 
Fig.~\ref{fig:disc_model_data}.
Viewed as a 3D distribution, this image is a delta function in the $v_z$ direction. It can of course be reconstructed using standard 2D tomography, but here the point is to see the outcome of reconstructing in 3D. If the problem was well constrained, we would recover an image that was strongly concentrated towards $v_z = 0$. The actual result, using a standard gaussian convolution default (with a FWHM of
\SI{200}{\km\per\s} as used in all other cases), is shown in projection in the top row of Fig.~\ref{fig:disc-proj}.
Here we are truly bitten by the missing information problem of 3D Doppler tomography! The fit to the data obtained from the image of the top row of
Fig.~\ref{fig:disc-proj} is perfectly acceptable, with a $\chi^2$ per data point of 1.00, identical to the fits obtained from the images shown in the other two rows of the figure (to be discussed below). The fit is not shown because it looks identical, albeit a little smoother, to the data of Fig.~\ref{fig:disc_model_data}.

The top row of Fig.~\ref{fig:disc-proj} is an explicit example of the disc--jet degeneracy problem discussed towards the end of section~\ref{sec:anything}. The two panels on the right show strong but entirely spurious spots of emission on the $v_z$ axis, symmetrically placed above and below the $v_x$--$v_y$ plane, which one could mistakenly interpret to be jets. These spots show "diffraction spike"-like cross-hairs that connect them back to the location of the accretion disc in the $v_x$--$v_y$ plane. These are in fact from the usual double cone seen in projection. The gaussian convolution default tends to favour the development of discrete spots within an image. In the 2D case, the data constraints are restrictive, and such spots genuine, but here it is possible for a spot on the $v_z$-axis to mimic a ring in the $v_x$--$v_y$ plane, and hence a strong and worrying artefact arises. 

This means one should be extremely wary before claiming evidence of out-of-plane motion from a given set of line profiles. Generated from data computed from an image with absolutely no motion out of the orbital plane, the top row of Fig.~\ref{fig:disc-proj} is instead dominated by emission at $v_z \neq 0$, and yet it delivers an excellent fit to the input line profiles.

What, if anything, can be done to address this problem? Just as alluded to in the disc--jet discussion, we need to inject some sort of extra "prior" information because the line profiles simply do not contain the information to rule out the emission pattern shown at the top of Fig.~\ref{fig:disc-proj}. Indeed, that pattern sprang directly from applying the inversion to the line profiles. In the case of disc mapping it was mentioned that the most axi-symmetric disc can be sought through the use of an azimuthal default image. A natural equivalent in the case of Doppler tomography is to search for the most compact image in the $v_z$ direction. We know that in most systems, much of the motion is indeed in the orbital plane, so the first question we might want to answer on a given target is whether there is any evidence for emission over a range of different $v_z$ values. If we try to force the image towards a delta function in the $v_z$-direction, and yet it refuses to approach that state, then we could be onto something.

I tested this idea as follows: the default at each new set of iterations was computed as a combination of the usual gaussian convolution of the previous image, plus a "squeezed" version constructed by replacing the image along the $v_z$ direction at every $v_x$--$v_y$ grid point by a narrow gaussian (of FWHM $=\SI{100}{\km\per\s}$ in the $v_z$ direction) centred at the flux-weighted mean $v_z$ value of the image at that grid-point. The squeezed version was convolved with a gaussian in the $v_x$ and $v_y$ directions. This couples neighbouring points in the $v_x$--$v_y$ grid and avoids the default becoming a set of independent columns in the $v_z$ direction. The two components were combined with a weighting factor or "pull" towards the squeezed image. Symbolically, the default was computed as
\begin{equation}
    J = (1-p)G(I) + p S(I).
\end{equation}
Here $G(I)$ is the standard gaussian-blurred default obtained from convolution with the 3D
gaussian of FWHM \SI{200}{\km\per\sec}. $S(I)$ is the squeezed default constructed
by first collapsing the image $I$ along the $v_z$ dimension, and then forming a 3D image by re-expanding along the $v_z$ direction as a gaussian, of fixed root mean square $\sigma$, centred upon the weighted centroid $\left<v_z\right>$ at each $v_x$--$v_y$ position. These steps are encapsulated by the
following expression
\begin{equation}
    \left(\int d \zeta\, I(v_x,v_y, \zeta)\right) 
    \frac{1}{\sqrt{2\pi} \sigma} \exp\left(-\frac{1}{2}\left(\frac{v_z-\left<v_z\right>}{\sigma}\right)^2 \right).
\end{equation}
This image was then blurred in $v_x$--$v_y$ to obtain the final pure squeezed default, $S(I)$. The pull factor $p$ is a number between $0$ and $1$ that controls the balance between the standard isotropic gaussian $G$ versus the pure squeezed default $S$, with the latter promoting a concentration of  emission at one location in the $v_z$ direction, if the data allow it.

The second row of Fig.~\ref{fig:disc-proj} shows the result of reconstructing with a 25\% pull given towards the squeezed image and 75\% towards the regular default ($p=0.25$). This is applied over many iterations, and it can be seen that it has allowed, as it was designed to do, a much greater intensity to build up in a narrow range in $v_z$. This is taken even further in the lowest row where a 50:50 weighting was used. Note that the location at $v_z = 0$ was not built in to the calculations, but emerged naturally as iterations proceeded. It is also worth emphasising once more that all three rows in Fig.~\ref{fig:disc-proj} represent identically good reconstructions in terms of the fit to the input data.
They all achieve $\chi^2/N = 1.00$.

For the largest 50\% pull factor, the $v_x$--$v_y$ projection (bottom-left of Fig.~\ref{fig:disc-proj}) approaches the appearance of the input image (left of Fig.~\ref{fig:disc_model_data}), whereas the $v_z$-axis "jets" are clearly visible in the other two rows. Here however the projections are a little misleading as the $v_z=0$ slices of the top two rows also look similar to the input image Fig.~\ref{fig:disc_model_data}, albeit with distinctly weakened disc components compared to the bottom row.

\subsection{A "polar"}

As outlined in the introduction, the need for three-dimensional imaging is most evident in the case of polars where there are very likely to be gas flows away from and back towards the orbital plane, so the final simulation is of a highly idealised "polar". I base this on results of 2D imaging of polars \cite{1997A&A...319..894S,1999MNRAS.304..145H,2000MNRAS.313..533S} which show evidence for a section of ballistic stream proceeding from the donor star, apparently little affected by the magnetic field, along with a magnetically-controlled structure associated with gas flowing down onto the white dwarf. The jump between these sections is quite sharp leading almost to a discontinuity in Doppler maps. It is the magnetically controlled section where one anticipates $v_z \neq 0$. I adopt an extremely simplistic representation of these structures with the emphasis being on generating easily understood input models as opposed to physically realistic ones. The latter is a challenge in any case, and it will be better to test these methods on real systems than to attempt to replicate them here.

The "polar" is therefore built from two finite length cylindrical structures in which the emission drops off as a gaussian with distance from the axis, over a speed scale of RMS $\sigma = \SI{30}{\km\per\s}$. The emissivity does not vary along the axis of the cylinder but sharply truncates at each end. One cylinder lies in the orbital plane parallel to $v_x$, and extends from $(-550,250,0)\,\si{\km\per\s}$ to $(+50,250,0)\,\si{\km\per\s}$, in primitive representation of the ballistic stream. The other is centred at $(0,-800,0)\,\si{\km\per\s}$
on the $v_y$ axis. Its axis has zero $v_x$ component but it is tilted at \ang{30} to the $v_x$--$v_y$ plane and has a total length of \SI{800}{\km\per\s}. This is to approximate a magnetically-controlled flow in which the gas first heads upwards with $v_z > 0$, reaches a maximum height out of the orbital plane ($v_z = 0$) before heading back towards it at accelerating speed with $v_z  < 0$. Thus this component clearly needs to intersect the $v_z = 0$ plane. This component in real systems lies in the negative $v_x$--$v_y$ quadrant; here I force it to have zero $v_x$ component to simplify the figures.

\begin{figure}
\centering
	\includegraphics[width=\columnwidth]{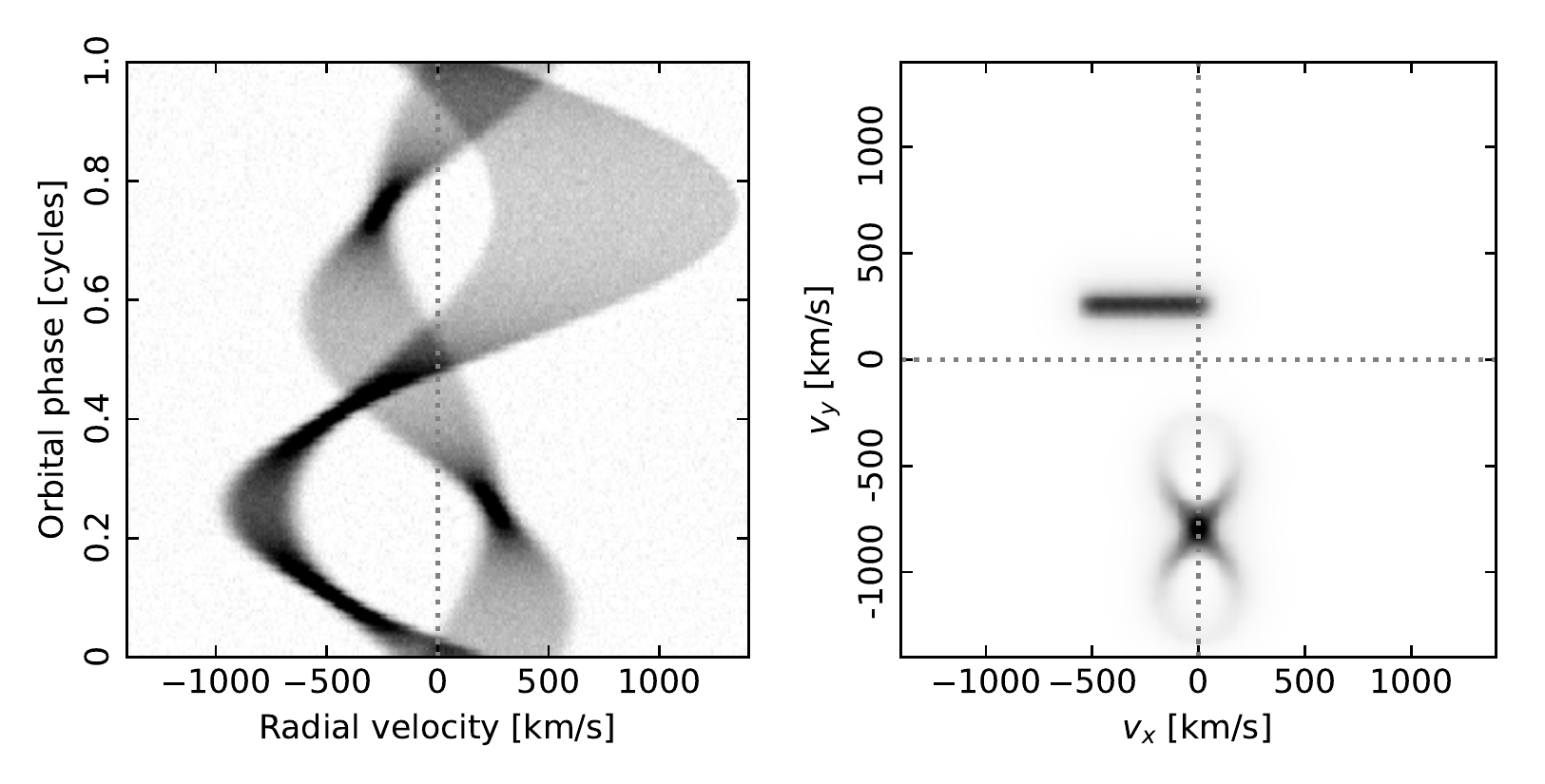}
    \caption{\emph{Left:} Data corresponding to a model "polar" consisting of two 
    cylindrical structures in 3D. One which is confined largely to the $v_x$--$v_y$ plane leads to the lower amplitude component. The other is tilted out of the $v_x$--$v_y$ plane causing the larger amplitude and less symmetric second component. \emph{Right:} a standard 2D map
    derived from the data. The map produces a poor fit to the data with
    $\chi^2$ per data point, $\chi^2/N =8$, set a little above the minimum it was able to reach to avoid corruption by noise.
    \label{fig:polar_data}}
\end{figure}

Following the usual procedure led to the data shown in Fig.~\ref{fig:polar_data}
where the data are plotted side-by-side with a standard 2D reconstruction.
The effect of the tilt out of the $v_x$--$v_y$ plane is seen in the data where the larger amplitude component does not have the symmetry of its in-plane counterpart. In the 2D reconstruction, the tilted component leads to another nice example of the double cone effect, being "in focus" at the point where it crosses $v_z = 0$, but spreading out into a ring pattern at either end. The resultant fit is poor with a $\chi^2$ per point $\chi^2/N = 8$ -- the data contain behaviour that cannot be captured by a 2D model.

\begin{figure}
\centering
	\includegraphics[width=0.9\columnwidth]{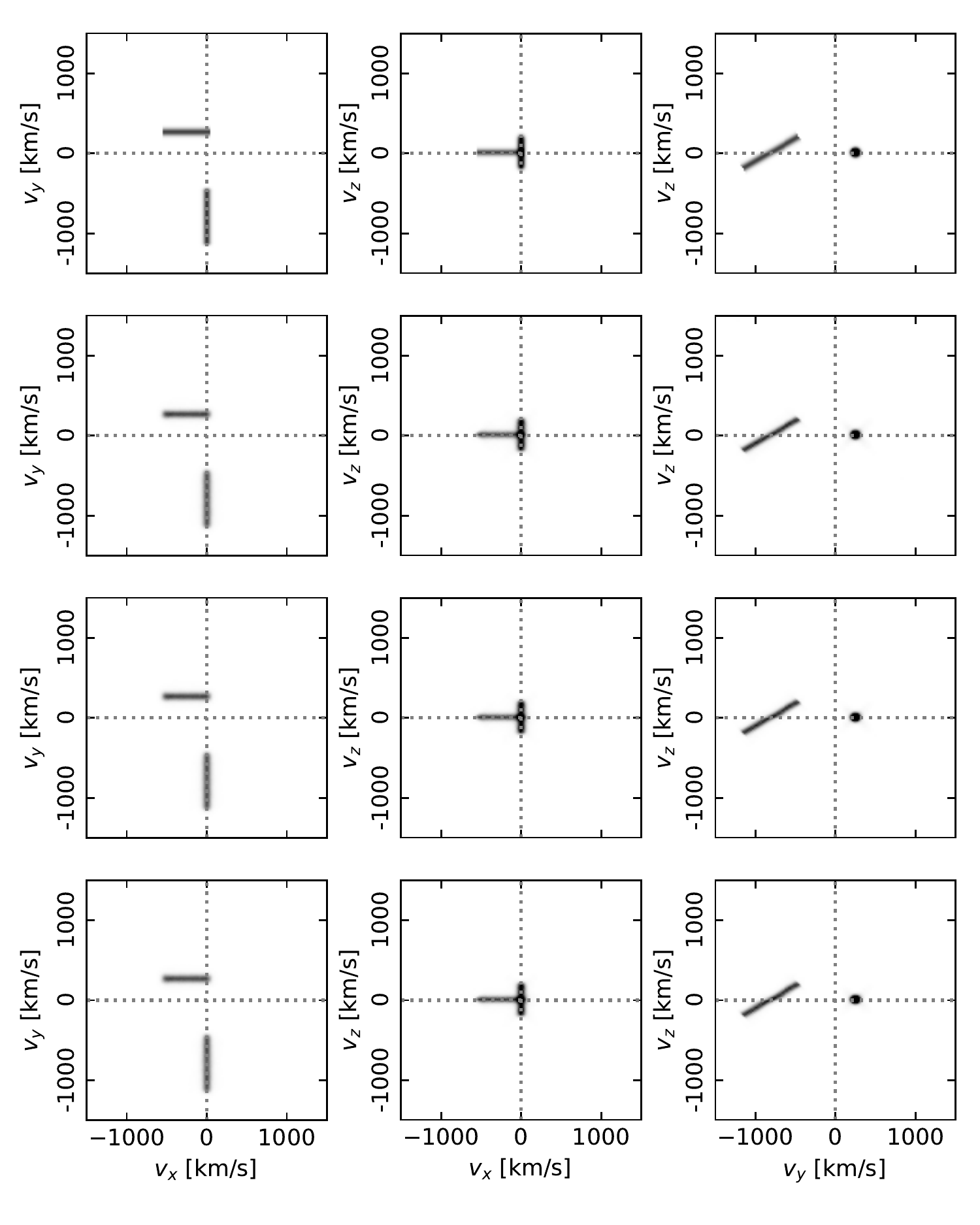}
    \caption{Top-to-bottom: the model and three reconstructions 
    based on the "polar" data of Fig.~\protect\ref{fig:polar_data}.
    All cases are shown in projection. The three reconstructions differ by the default used. For the second row a simple gaussian
    default was used while the third and fourth rows used a 25\% and 50\% squeezed default. Each reconstruction fits the data well with
    $\chi^2/N = 1$.
    \label{fig:polar-proj}}
\end{figure}

\begin{figure}
\centering
	\includegraphics[width=0.9\columnwidth]{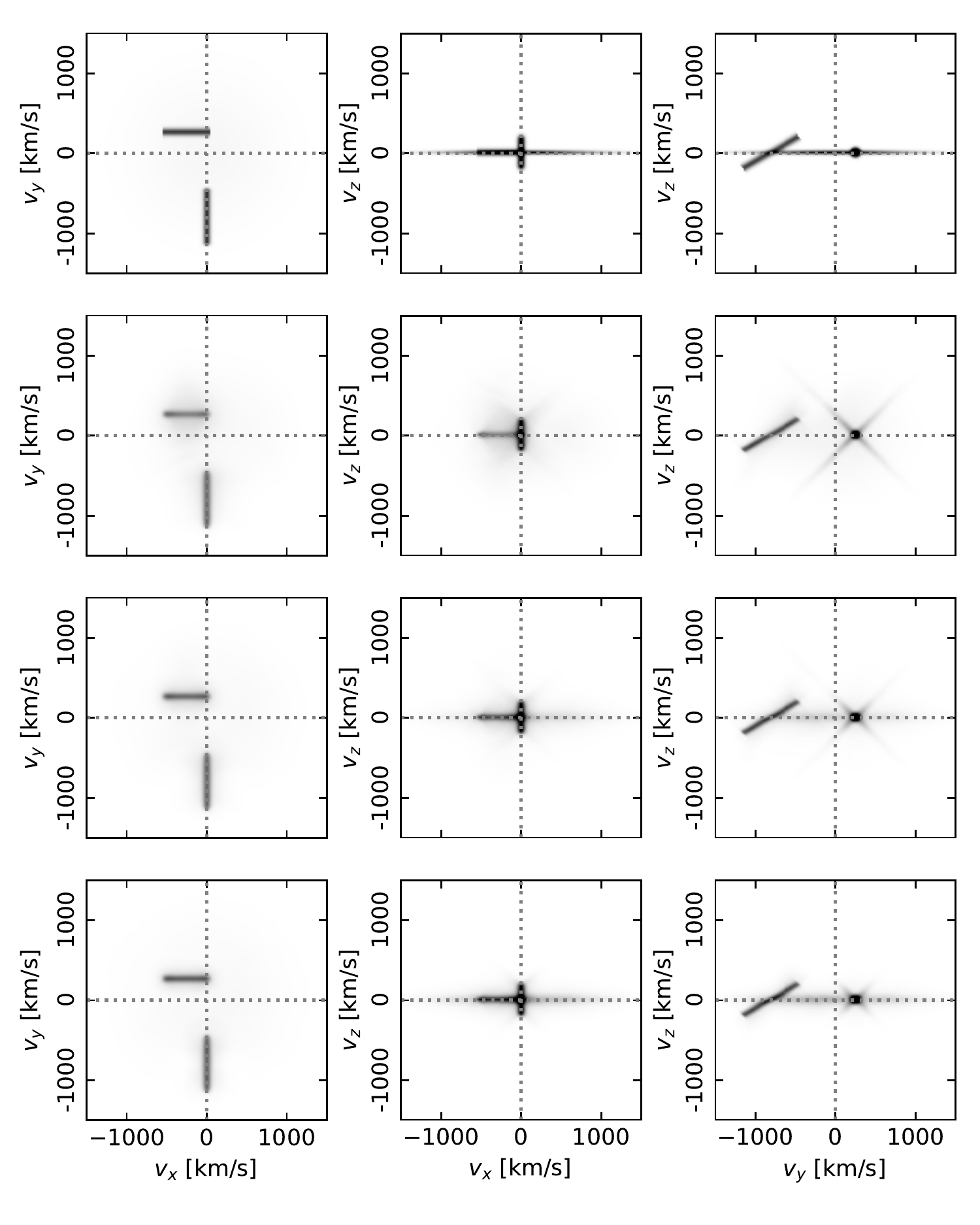}
    \caption{The model and three reconstructions 
    based upon the same "polar" model used for Fig.~\ref{fig:polar-proj}
    but with the addition of a broad gaussian emission component spread out in the $v_x$--$v_y$ plane, and best seen in the edge-on projections in the two top-right panels. See the caption of Fig.~\ref{fig:polar-proj} for an explanation of the row order.
    \label{fig:polar-proj2}}
\end{figure}

Starting from these data, three 3D reconstructions were performed using a plain gaussian default, and
squeezed defaults with 25\% and 50\% pull factors, as were used in the disc reconstructions of Fig.~\ref{fig:disc-proj}. These reconstructions, along with the original model, are shown in projection in Fig.~\ref{fig:polar-proj}.
In all cases $\chi^2/N = 1$ was reached.
The reconstructions in this admittedly simplified scenario are a return to low level artefacts and the differences between the reconstructions and the model appear to be relatively minor. This may again be related to the compact nature of the structures and positivity constraints. Therefore as a final test, a double gaussian disc was added of the form
\begin{equation}
    \exp\left(-\frac{v_z^2}{2\sigma_z^2}\right)
    \exp\left(-\frac{v_x^2+v_y^2}{2\sigma_{xy}^2}\right),
\end{equation}
with $\sigma_z = \SI{20}{\km\per\s}$ and $\sigma_{xy} = \SI{600}{\km\per\s}$. This was not intended to model anything specific, but just to be representative of the sort of additional emission components of unclear origin one sometimes sees in real systems. The above component is strongest in a narrow band near $v_z = 0$, but broadly spread over the $v_x$--$v_y$ plane. Its total flux was scaled to match the total flux from the cylindrical components. The model and reconstructions, exactly equivalent to Fig.~\ref{fig:polar-proj} for the cylindrical components on their own, are displayed in Fig.~\ref{fig:polar-proj2}. 

As might have been expected given the similarities with the disc simulation, some clear artefacts now appear. In the model (top row), the presence of the extra gaussian slab is clear in the two 
side-on projections at the top right of the plot. However, it is not seen at all in the gaussian convolution reconstruction (second row).  Instead, clear bi-conical artefacts appear. They are particularly evident in the right-most panel where the ballistic stream is seen end on. The two lower squeezed default rows show partial recovery of the gaussian slab,  and
a reduction in strength of the bi-conical artefacts, but they also reveal a new effect in the form of slight kinks in the sloping magnetic flow component. This is a result of the gaussian slab's contribution towards the weighted mean value of $v_z$ during the computation of the squeezed default. 

These experiments reveal a complex interplay between different structures, but at the same time, they are modestly encouraging of the view that there may be something to be learned from the application of 3D imaging to real systems, in spite of the missing information.

\section{Discussion}

Tomography in 3D is hard to pin down. There is for sure a problem with missing information, seen most clearly in Fourier terms, which makes it qualitatively different from the 2D case. On the other hand, sometimes at least, qualitatively correct reconstructions appear to emerge. However, Figs~\ref{fig:single-mod} and \ref{fig:disc-proj} warn of the potential for peril in the pursuit of 3D images. In Fig.~\ref{fig:disc-proj}, data generated from a 2D disc image, with emission precisely confined to $v_z = 0$, results in a 3D map in which the bulk of the emission lies at $|v_z| > \SI{500}{\km\per\s}$. The fit to the data from this manifestly corrupted image is fine, with $\chi^2/N = 1$, and, moreover, equally good fits to the data can be obtained with extremely different looking images (lower two rows). This result is compelling given that the data were simulated with higher than typical signal-to-noise; in a more realistic case, there would be significantly more freedom still. \emph{There is nothing in the data} to distinguish between the three reconstructions shown row-by-row in Fig.~\ref{fig:disc-proj}.

The potential for such artefacts depends very much upon the
emissivity distribution itself. If emission is concentrated into a few relatively compact structures, useful reconstructions can emerge. This may explain the results of \citet{2014ARep...58..881K} who were able to recover relatively simple input emission distributions for polars that were created from MHD simulations. However, the example of disc emission shows the danger of the disc-vs-jet degeneracy and, for instance, suggests that the jet-like outflow seen in U~CrB by \citet{2009ApJ...690.1730A} could be spurious. As far as I can tell, the potential impacts of such degeneracy have not been recognised before, as, had they been, it is perhaps unlikely that \citet{2018ARep...62...89A} would have said that artefacts were reduced to a negligible level using the CLEAN algorithm. No algorithm, maximum entropy or CLEAN or any other, can make up for the missing Fourier components in the 3D case, and this applies even if data could be acquired with infinite signal-to-noise and resolution, and at all binary phases. We need other assumptions -- positivity for example -- to help us out.

For relatively simple structures, positivity seems to get us a long way, but may fail in more complex ones such as the disc emission example. A difficulty in practice will be knowing when such failures have occurred, as there are no "input models" when it comes to real data. It might therefore always be wise to start 3D tomography analyses with the bar set lower, aiming first of all to answer whether the data provide any compelling evidence for a finite spread over the off-plane component, $v_z$. I presented one possible approach using what I term a "squeezed default" where the aim is to attempt to concentrate the emission in the $v_z$ direction, but it might need modification according to the system under investigation. Even this low level target may prove difficult to achieve in practice, as very often 2D images struggle to fit data of good signal-to-noise and a 3D image is almost bound to find a better solution in terms of $\chi^2$ in such cases. It then may become a matter of subjective judgement whether one believes the results, as motion in $v_z$ is not the only way in which line profiles can become impossible to match with 2D images.

At the very least, if applying the maximum entropy method of this paper, the effect of the choice of default upon a reconstruction at fixed $\chi^2$ should always be investigated. In the case of the elementary polar of Fig.~\ref{fig:polar-proj}, it made little difference whether a homogeneous gaussian default or one squeezed towards constant $v_z$ was used, whereas the images of Figs~\ref{fig:disc-proj} and \ref{fig:polar-proj2} were significantly affected by this choice. The same effects for real data might offer some assurance that features were real in the first instance, or potentially false in the second.

\section{Summary and Conclusions}

In this paper I have considered to what extent line profiles from binary systems can be used to uncover their emissivity distribution in 3D velocity space, i.e. over $v_x$, $v_y$ and $v_z$, adding motion parallel to the orbital axis of the binary as well as the motion parallel to the orbital plane that is accounted for in the standard 2D version of Doppler tomography. The problem is simplest to analyse in terms of Fourier transforms. Line profiles directly constrain the 3D Fourier transform of the 3D emissivity image in velocity coordinates, but only on a 2D surface that has the shape of a double cone in Fourier \kvec-space aligned with the $k_{vz}$ axis, and centred upon the origin. This very partial information rules out a well-constrained inversion comparable to the usual 2D imaging case where the full 2D Fourier transform is potentially obtainable from data.

Some information on the 3D emissivity distribution is nevertheless contained within binary star line profiles. A straightforward extension of the 2D imaging method developed by \citet{1988MNRAS.235..269M} was implemented and applied to a number of test cases. The fidelity of the reconstructed images was found to depend very much upon the form of the test image. In some cases the essential form of the input image was returned to a large extent, whereas some others were corrupted by artefacts. To combat the latter problem, a method was presented to steer the reconstruction towards an essentially 2D distribution in an attempt to answer the more elementary question of whether the data contain any evidence for 3D motion.

These results allow cautious hope that there might be some scope to apply 3D tomography to real data, but great care is needed in the interpretation of results from such an exercise. Three dimensional tomography is very different from its two dimensional relation, and the experience built up from decades of application of the latter may not be a useful guide. Any such study is likely to require an element of simulations to assess the possible impact of reconstruction artefacts. It will be easy to obtain results, but much harder to judge their veracity.

\section*{Acknowledgements}

I thank Geoff Daniell, Mark Cropper, Keith Horne and Danny Steeghs for conversations about Doppler tomography over the years. This research was supported by a Leverhulme Research Fellowship. I thank the anonymous referee for their careful reading of the manuscript and for helpful suggestions.

\section*{Data Availability}

This paper is based upon software developed by the author. The code is published on the
{\tt github} platform\footnote{https://github.com/trmrsh/trm-doppler}.
Key features of the code are described in appendix~\ref{app:dcode}.



\bibliographystyle{mnras}
\bibliography{dopp3d} 

\begin{thebibliography}{}
\makeatletter
\relax
\def\mn@urlcharsother{\let\do\@makeother \do\$\do\&\do\#\do\^\do\_\do\%\do\~}
\def\mn@doi{\begingroup\mn@urlcharsother \@ifnextchar [ {\mn@doi@}
  {\mn@doi@[]}}
\def\mn@doi@[#1]#2{\def\@tempa{#1}\ifx\@tempa\@empty \href
  {http://dx.doi.org/#2} {doi:#2}\else \href {http://dx.doi.org/#2} {#1}\fi
  \endgroup}
\def\mn@eprint#1#2{\mn@eprint@#1:#2::\@nil}
\def\mn@eprint@arXiv#1{\href {http://arxiv.org/abs/#1} {{\tt arXiv:#1}}}
\def\mn@eprint@dblp#1{\href {http://dblp.uni-trier.de/rec/bibtex/#1.xml}
  {dblp:#1}}
\def\mn@eprint@#1:#2:#3:#4\@nil{\def\@tempa {#1}\def\@tempb {#2}\def\@tempc
  {#3}\ifx \@tempc \@empty \let \@tempc \@tempb \let \@tempb \@tempa \fi \ifx
  \@tempb \@empty \def\@tempb {arXiv}\fi \@ifundefined
  {mn@eprint@\@tempb}{\@tempb:\@tempc}{\expandafter \expandafter \csname
  mn@eprint@\@tempb\endcsname \expandafter{\@tempc}}}

\bibitem[\protect\citeauthoryear{{Agafonov}, {Richards}  \&
  {Sharova}}{{Agafonov} et~al.}{2006}]{2006ApJ...652.1547A}
{Agafonov} M.,  {Richards} M.,   {Sharova} O.,  2006, \mn@doi [\apj]
  {10.1086/508484}, \href
  {https://ui.adsabs.harvard.edu/abs/2006ApJ...652.1547A} {652, 1547}

\bibitem[\protect\citeauthoryear{{Agafonov}, {Sharova}  \&
  {Richards}}{{Agafonov} et~al.}{2009}]{2009ApJ...690.1730A}
{Agafonov} M.~I.,  {Sharova} O.~I.,   {Richards} M.~T.,  2009, \mn@doi [\apj]
  {10.1088/0004-637X/690/2/1730}, \href
  {https://ui.adsabs.harvard.edu/abs/2009ApJ...690.1730A} {690, 1730}

\bibitem[\protect\citeauthoryear{{Agafonov}, {Karitskaya}, {Sharova},
  {Bochkarev}, {Zharikov}, {Butenko}, {Bondar'}  \& {Sidorov}}{{Agafonov}
  et~al.}{2018}]{2018ARep...62...89A}
{Agafonov} M.~I.,  {Karitskaya} E.~A.,  {Sharova} O.~I.,  {Bochkarev} N.~G.,
  {Zharikov} S.~V.,  {Butenko} G.~Z.,  {Bondar'} A.~V.,   {Sidorov} M.~Y.,
  2018, \mn@doi [Astronomy Reports] {10.1134/S1063772918020026}, \href
  {https://ui.adsabs.harvard.edu/abs/2018ARep...62...89A} {62, 89}

\bibitem[\protect\citeauthoryear{{Astropy Collaboration} et~al.,}{{Astropy
  Collaboration} et~al.}{2013}]{2013A&A...558A..33A}
{Astropy Collaboration} et~al., 2013, \mn@doi [\aap]
  {10.1051/0004-6361/201322068}, \href
  {https://ui.adsabs.harvard.edu/abs/2013A&A...558A..33A} {558, A33}

\bibitem[\protect\citeauthoryear{{Astropy Collaboration} et~al.,}{{Astropy
  Collaboration} et~al.}{2018}]{2018AJ....156..123A}
{Astropy Collaboration} et~al., 2018, \mn@doi [\aj] {10.3847/1538-3881/aabc4f},
  \href {https://ui.adsabs.harvard.edu/abs/2018AJ....156..123A} {156, 123}

\bibitem[\protect\citeauthoryear{{Heerlein}, {Horne}  \& {Schwope}}{{Heerlein}
  et~al.}{1999}]{1999MNRAS.304..145H}
{Heerlein} C.,  {Horne} K.,   {Schwope} A.~D.,  1999, \mn@doi [\mnras]
  {10.1046/j.1365-8711.1999.02311.x}, \href
  {https://ui.adsabs.harvard.edu/abs/1999MNRAS.304..145H} {304, 145}

\bibitem[\protect\citeauthoryear{{Horne}}{{Horne}}{1985}]{1985MNRAS.213..129H}
{Horne} K.,  1985, \mn@doi [\mnras] {10.1093/mnras/213.2.129}, \href
  {https://ui.adsabs.harvard.edu/abs/1985MNRAS.213..129H} {213, 129}

\bibitem[\protect\citeauthoryear{{Horne} \& {Marsh}}{{Horne} \&
  {Marsh}}{1986a}]{1986MNRAS.218..761H}
{Horne} K.,  {Marsh} T.~R.,  1986a, \mn@doi [\mnras] {10.1093/mnras/218.4.761},
  \href {https://ui.adsabs.harvard.edu/abs/1986MNRAS.218..761H} {218, 761}

\bibitem[\protect\citeauthoryear{{Horne} \& {Marsh}}{{Horne} \&
  {Marsh}}{1986b}]{1986LNP...266....1H}
{Horne} K.,  {Marsh} T.~R.,  1986b, {Indirect Imaging of Accretion Disks in
  Binaries}.
p.~1, \mn@doi{10.1007/3-540-17195-9\_1}

\bibitem[\protect\citeauthoryear{{Huang}}{{Huang}}{1972}]{1972ApJ...171..549H}
{Huang} S.-S.,  1972, \mn@doi [\apj] {10.1086/151309}, \href
  {https://ui.adsabs.harvard.edu/abs/1972ApJ...171..549H} {171, 549}

\bibitem[\protect\citeauthoryear{{Kononov}, {Agafonov}, {Sharova}, {Bisikalo},
  {Zhilkin}  \& {Sidorov}}{{Kononov} et~al.}{2014}]{2014ARep...58..881K}
{Kononov} D.~A.,  {Agafonov} M.~I.,  {Sharova} O.~I.,  {Bisikalo} D.~V.,
  {Zhilkin} A.~G.,   {Sidorov} M.~Y.,  2014, \mn@doi [Astronomy Reports]
  {10.1134/S1063772914120063}, \href
  {https://ui.adsabs.harvard.edu/abs/2014ARep...58..881K} {58, 881}

\bibitem[\protect\citeauthoryear{{Kraft}, {Mathews}  \& {Greenstein}}{{Kraft}
  et~al.}{1962}]{1962ApJ...136..312K}
{Kraft} R.~P.,  {Mathews} J.,   {Greenstein} J.~L.,  1962, \mn@doi [\apj]
  {10.1086/147381}, \href
  {https://ui.adsabs.harvard.edu/abs/1962ApJ...136..312K} {136, 312}

\bibitem[\protect\citeauthoryear{{Manser} et~al.,}{{Manser}
  et~al.}{2016}]{2016MNRAS.455.4467M}
{Manser} C.~J.,  et~al., 2016, \mn@doi [\mnras] {10.1093/mnras/stv2603}, \href
  {https://ui.adsabs.harvard.edu/abs/2016MNRAS.455.4467M} {455, 4467}

\bibitem[\protect\citeauthoryear{{Marsh}}{{Marsh}}{2001}]{2001LNP...573....1M}
{Marsh} T.~R.,  2001, {Doppler Tomography}.
p.~1

\bibitem[\protect\citeauthoryear{{Marsh}}{{Marsh}}{2005}]{2005Ap&SS.296..403M}
{Marsh} T.~R.,  2005, \mn@doi [\apss] {10.1007/s10509-005-4859-3}, \href
  {https://ui.adsabs.harvard.edu/abs/2005Ap&SS.296..403M} {296, 403}

\bibitem[\protect\citeauthoryear{{Marsh} \& {Horne}}{{Marsh} \&
  {Horne}}{1988}]{1988MNRAS.235..269M}
{Marsh} T.~R.,  {Horne} K.,  1988, \mn@doi [\mnras] {10.1093/mnras/235.1.269},
  \href {https://ui.adsabs.harvard.edu/abs/1988MNRAS.235..269M} {235, 269}

\bibitem[\protect\citeauthoryear{{Ramachandran} \& {Varoquaux}}{{Ramachandran}
  \& {Varoquaux}}{2011}]{2011CSE....13b..40R}
{Ramachandran} P.,  {Varoquaux} G.,  2011, \mn@doi [Computing in Science and
  Engineering] {10.1109/MCSE.2011.35}, \href
  {https://ui.adsabs.harvard.edu/abs/2011CSE....13b..40R} {13, 40}

\bibitem[\protect\citeauthoryear{{Richards}, {Agafonov}  \&
  {Sharova}}{{Richards} et~al.}{2012}]{2012ApJ...760....8R}
{Richards} M.~T.,  {Agafonov} M.~I.,   {Sharova} O.~I.,  2012, \mn@doi [\apj]
  {10.1088/0004-637X/760/1/8}, \href
  {https://ui.adsabs.harvard.edu/abs/2012ApJ...760....8R} {760, 8}

\bibitem[\protect\citeauthoryear{{Schwope}, {Mantel}  \& {Horne}}{{Schwope}
  et~al.}{1997}]{1997A&A...319..894S}
{Schwope} A.~D.,  {Mantel} K.~H.,   {Horne} K.,  1997, \aap, \href
  {https://ui.adsabs.harvard.edu/abs/1997A&A...319..894S} {319, 894}

\bibitem[\protect\citeauthoryear{{Schwope}, {Catal{\'a}n}, {Beuermann},
  {Metzner}, {Smith}  \& {Steeghs}}{{Schwope}
  et~al.}{2000}]{2000MNRAS.313..533S}
{Schwope} A.~D.,  {Catal{\'a}n} M.~S.,  {Beuermann} K.,  {Metzner} A.,  {Smith}
  R.~C.,   {Steeghs} D.,  2000, \mn@doi [\mnras]
  {10.1046/j.1365-8711.2000.03240.x}, \href
  {https://ui.adsabs.harvard.edu/abs/2000MNRAS.313..533S} {313, 533}

\bibitem[\protect\citeauthoryear{{Skilling} \& {Bryan}}{{Skilling} \&
  {Bryan}}{1984}]{1984MNRAS.211..111S}
{Skilling} J.,  {Bryan} R.~K.,  1984, \mn@doi [\mnras]
  {10.1093/mnras/211.1.111}, \href
  {https://ui.adsabs.harvard.edu/abs/1984MNRAS.211..111S} {211, 111}

\bibitem[\protect\citeauthoryear{{Smak}}{{Smak}}{1981}]{1981AcA....31..395S}
{Smak} J.,  1981, \actaa, \href
  {https://ui.adsabs.harvard.edu/abs/1981AcA....31..395S} {31, 395}

\bibitem[\protect\citeauthoryear{{Steeghs}}{{Steeghs}}{2003}]{2003MNRAS.344..448S}
{Steeghs} D.,  2003, \mn@doi [\mnras] {10.1046/j.1365-8711.2003.06917.x}, \href
  {https://ui.adsabs.harvard.edu/abs/2003MNRAS.344..448S} {344, 448}

\bibitem[\protect\citeauthoryear{{Tapia}}{{Tapia}}{1977}]{1977ApJ...212L.125T}
{Tapia} S.,  1977, \mn@doi [\apjl] {10.1086/182390}, \href
  {https://ui.adsabs.harvard.edu/abs/1977ApJ...212L.125T} {212, L125}

\makeatother
\end{thebibliography}




\appendix

\section{Python-based Doppler tomography code}

\label{app:dcode}

The computations in this paper were carried out using a mixed {\tt Python} and {\tt C++}-based implementation of the method presented by \citet{1988MNRAS.235..269M}. {\tt Python} here acts as the interface between the data and Doppler images which are stored in FITS files and the computations which are devolved to {\tt C++} subroutines. This combines the ease of use and flexibility of {\tt Python}, for instance to access the {\tt astropy} software suite \citep{2013A&A...558A..33A,2018AJ....156..123A} with its routines for handling FITS-format data, with the speed of {\tt C}-code. Thus the image and data files are read into {\tt numpy} arrays at the {\tt Python} top level and then sent to {\tt C++}-routines. The arrays returned are then dealt with by {\tt Python}. There is some complexity in the interface code itself, but there should be no need for users to engage with this aspect; at the user level, the code appears as a set of {\tt Python} methods and classes, and associated scripts can be used to operate in command line mode, which is the usual anticipated usage. Together these form a {\tt Python} module called {\tt trm.doppler}.

The details are best uncovered by examining the code, but, to give the gist of it, the heart of the software lies in the subroutine that implements Eq.~\ref{eq:matrix}.
This contains seven nested loops which iterate in turn over the following: (1) each data set; (2) each image; (3) each spectrum;
(4) each sub-spectrum within a spectrum to simulate finite length exposures; (5), (6) and (7), the $v_z$, $v_y$ and $v_x$ axes of the image. A OpenMP (omp or open multi-processing) parallelisation directive is applied prior to loop (3) over the spectra since the effort required per spectrum from a given data set is very similar and well balanced. The innermost three loops which iterate over all elements of the images are written to operate as efficiently as possible. Intermediate finely-spaced buffers are used to implement blurring operations to represent finite instrumental resolution, which is allowed to vary from data set to data set. The blurring itself is implemented with Fourier Transforms. Fourier transforms are also used for the blurring operations often needed during default computation.

As a piece of software that is likely to undergo changes in the future, there is no point in attempting too detailed a description here, so I confine myself to a description of its key features, as these might prove the best guide for those wondering whether to try out the software. These are as follows:

\emph{FITS-based data and image model:} Both the data and the images
have a FITS-based format, each with multiple header data units (HDUs). This is a move away from the previous {\tt F77}-based code which relied upon libraries developed for the UK's {\tt STARLINK} project \citep{1988MNRAS.235..269M}. The FITS routines of the {\tt astropy} project are used to read and write these files. The first step in any usage of the software is conversion of one's data into the required 
FITS input data structure.

\emph{Multiple independent data sets as inputs:} One can reconstruct images using more than one source of data. 
For instance, one could use data covering H$\alpha$, say, from different telescopes and instruments, each with a different resolution, coverage and sampling. Each set of data appears as a set of HDUs containing fluxes, errors, wavelengths and phases or times. Within such a set, the fluxes are stored as 2D data, but if other data are taken with an incompatible sampling and resolution, then another set of HDUs representing them can be added.
The {\tt F77}-code only operated on single homogeneous data sets taken with a single instrument and telescope, with a fixed configuration throughout the run. This feature is of particular use in long-running monitoring experiments where it can be close to impossible to ensure uniform instrumentation throughout. A good example of just
this is the Doppler imaging study published by \citet{2016MNRAS.455.4467M} which combined data from multiple instruments taken over more than a decade to obtain an image of a slowly precessing debris disc around an isolated white dwarf.

\emph{Raw wavelength scale:} To use the {\tt F77}-code, one had to re-bin one's data to have a single scale for all spectra that was uniform in terms of radial velocity step from pixel to pixel. This is no longer required, as the new code works with arbitrary wavelength scales, although they are assumed to vary smoothly.

\emph{Multiple and blended lines:} An arbitrary number of atomic lines can be imaged at the same time. They can be overlapping in the data. It is also possible to define one image as representative of more than one line in the data. For instance, one may believe that all Balmer lines are essentially the same except for a scaling factor. Then one can link all the Balmer lines to a single image, along with appropriate scaling factors (that can themselves be optimised as part of the optimisation).

\emph{Finite exposure times:} for short period, faint systems, it can be difficult to take exposures that are not a significant fraction of the orbital period. This can be allowed for by defining setups in which each spectrum is computed by sub-dividing the exposure and trapezoidally-averaging across the exposure. The effect of this can be to reduce the amount of azimuthal smearing in images, although it is usually only partially successful in practice.

\emph{Speed:} The code operates fast. Standard 2D imaging operations take much less than a second, and a complete set of iterations in such cases may only require a few seconds. The OMP parallelisation allows it to utilise multi-core machines efficiently, which is of particular importance in the 3D case.

\emph{Modulation mapping:} \citet{2003MNRAS.344..448S} introduced an extension to Doppler tomography to account for the common issue of components that vary in flux with orbital phase. Such variations are not accommodated by "classical" Doppler tomography \citep{2001LNP...573....1M}. The new code allows for this with the addition of extra component whose contribution is added in after multiplication by sine and cosine terms, as explained in further detail by \citet{2003MNRAS.344..448S}.

\emph{Negative fluxes:} As shown in this paper, positivity can greatly suppress artefacts in otherwise poorly constrained reconstructions. Nevertheless, it is very common in practice to encounter data which could be much better fit if the image could become at least partially negative. This is common for instance when the line emission lies on top of absorption, perhaps from a white dwarf. Another common case occurs in high inclination systems where the disc can absorb light from the white dwarf leading to a deep central cores to the lines. While negative fluxes are not physical, allowing for them may enable the fit to have an easier time fitting the data, revealing features that can be missed if positivity is rigidly adhered to. The new code allows for this in rather the same way as for modulation mapping by the introduction of additional image components, which although they are individually entirely positive
to allow the computation of the entropy, contribute negatively to the line profiles. Given the results of this paper, it is likely that this option should only ever be used for 2D imaging.

\emph{2D and 3D:} Last but not least, obviously the code can handle both conventional 2D as well as 3D Doppler tomography, but, as the paper should also be clear, the 3D case will always require special care over the reality of features. It may however be of some value simply as a more flexible fitting tool than the 2D version.


\bsp	
\label{lastpage}
\end{document}